\documentclass[onecolumn,superscriptaddress,prc,nofootinbib,reprint]{revtex4-1}

\usepackage{amsmath}
\usepackage{amssymb}
\usepackage{color}
\usepackage{graphicx}

\newcommand{\be}{\begin{equation}}
\newcommand{\ee}{\end{equation}}
\newcommand{\ba}{\begin{eqnarray}}
\newcommand{\ea}{\end{eqnarray}}
\newcommand{\nn}{\nonumber}
\newcommand{\etaeff}{\left\langle \eta/s \right\rangle_{\textrm{eff}}}
\newcommand{\zetaeff}{\left\langle \zeta/s \right\rangle_{\textrm{eff}}}

\begin{document}

\title{Effective viscosities in a hydrodynamically expanding boost-invariant QCD plasma}
\author{Jean-Fran\c{c}ois Paquet}
\author{Steffen A. Bass}
\affiliation{Department of Physics, Duke University, Durham, NC 27708, USA}
\date{\today}

\begin{abstract}
	\begin{description}
		\item[Background] 
		The near-equilibrium properties of a QCD plasma can be encoded into transport coefficients such as bulk and shear viscosity. In QCD, the ratio of these transport coefficients to entropy density, $\zeta/s$ and $\eta/s$, depends non-trivially on the plasma's temperature. This is unlike in conformal systems where they take constant values such as $\eta/s=1/(4\pi)$.
		\item[Purpose]
		In this work, we show that in a 0+1D boost-invariant fluid with no transverse expansion, a temperature-dependent $\zeta/s(T)$ or $\eta/s(T)$ can be %
		described by an equivalent \emph{effective} viscosity $\zetaeff$ or $\etaeff$.
		This effective viscosity combines the actual temperature-dependent $\zeta/s(T)$ or $\eta/s(T)$ with the temperature profile of the fluid.
		We extend the concept of effective viscosity in systems with transverse expansion, and discuss how effective viscosities can be used to identify families of $\zeta/s(T)$ and $\eta/s(T)$ that lead to similar hydrodynamic evolution.
		\item[Method] 
		The Navier-Stokes relativistic hydrodynamic equations are used to provide a first definition of effective viscosity, in 0+1D and 1+1D. In the 0+1D case, the analysis is extended to Israel-Stewart-type second-order hydrodynamics to clarify the effect of higher-order hydrodynamic corrections on the effective viscosity.
		\item[Results]
		In a boost-invariant fluid with no transverse expansion (0+1D), the effective viscosity is expressed as a simple integral of $\zeta/s(T)$ or $\eta/s(T)$ over temperature, with a weight determined by the speed of sound of the fluid. The result is general for any equation of state with a moderate temperature dependence of the speed of sound, including the QCD equation of state.
		This definition of effective viscosity can be used to identify infinite families of $\zeta/s(T)$ or $\eta/s(T)$ that produce essentially indistinguishable temperature profiles.
		In a boost-invariant cylindrical system (1+1D), a similar definition of effective viscosity is obtained in terms of characteristic trajectories in time and transverse direction. This leads to an infinite number of constraints on an infinite functional space for $\zeta/s(T)$ and $\eta/s(T)$. Realistic examples are presented using a finite number of constraints on a finite functional space.
		\item[Conclusions] 
		The definition of effective viscosity in a 0+1D system  clarifies how infinite families of $\zeta/s(T)$ and $\eta/s(T)$ can result in nearly identical hydrodynamic temperature profiles.
		By extending the study to a boost-invariant cylindrical (1+1D) fluid, we identify an approximate but more general definition of effective viscosity that highlight the potential and limits of the concept of effective viscosity in fluids with limited symmetries.
	\end{description}
\end{abstract}

\maketitle

\section{Introduction}

An important objective of the heavy ion program at the Relativistic Heavy ion Collider (RHIC) and the Large Hadron Collider (LHC) is to study the many-body (near-equilibrium) properties of Quantum Chromodynamics (QCD).\footnote{See Ref.~\cite{Busza:2018rrf} for a recent overview of the goals and open questions of heavy ion physics.} It is generally agreed that the shear viscosity to entropy density ratio of QCD is of order $\eta/s\sim 0.1$ in the temperature range $T\sim 150-500$~MeV probed in heavy ion collisions. More precise phenomenological constraints on the temperature dependence of $\eta/s$ are still under investigations (see for example Ref.~\cite{Bernhard:2019bmu}). The same is true for the bulk viscosity of QCD, which is being investigated in parallel. Advances are also being made from the theoretical side with a variety of approaches (see Ref.~\cite{Arnold:2006fz,Moore:2008ws,NoronhaHostler:2008ju,Laine:2014hba,Rose:2017bjz,Ghiglieri:2018dib,Czajka:2018bod} for example). 

Measurements from heavy ion collisions provide indirect phenomenological constraints on the viscosities of QCD. The link between experimental data and the viscosities is a multistage model~\cite{Gale:2013da,deSouza:2015ena} which describes the different successive phases of a heavy ion collision until only colorless particles with negligible interactions remain. The core of this multistage model is viscous relativistic hydrodynamics, which is used to describe the space-time evolution of the deconfined QCD plasma. This hydrodynamic description of heavy ion collisions is restricted to space-time regions of high energy density which achieve near local equilibrium; other models are used in other regions of the collisions. The viscosity of QCD affects the space-time expansion of the plasma, which is later reflected in the momentum distribution of the final colorless particles. The multistage model's prediction for the momentum distribution of colorless particle is then compared with experimental measurements, providing constraints on the transport coefficients of QCD.

The ratios of QCD's shear and bulk viscosities to entropy density, $\eta/s$ and $\zeta/s$, are unquestionably temperature-dependent. How significant this temperature dependence is, in the range of temperatures probed in heavy ion collisions, is still under investigation. At least for shear viscosity, it is still common to assume the temperature dependence to be modest; sufficiently modest to be well approximated by a constant ``effective'' shear viscosity, $\etaeff$. This effective viscosity is understood to be some average of the temperature dependent $\eta/s(T)$ over the temperature profile of the plasma.

Providing a general definition of this effective viscosity is challenging. In this work, we use systems with strong symmetries as an introductory approach to the concept of effective viscosity. We begin with the so-called Bjorken symmetries~\cite{Bjorken:1982qr}, a 0+1D boost-invariant fluid with no transverse dynamics. We follow with a 1+1D boost-invariant system with cylindrically-symmetric transverse expansion. Boost-invariance holds to a good approximation in the midrapidity region of heavy ion collisions. Cylindrically-symmetric transverse expansion may be a reasonable approximation for head-on heavy ion collisions. Neither will apply directly to a typical heavy ion collisions, yet they can capture many features of these collisions and provide useful guidance.

We begin with 0+1D Bjorken first-order (Navier-Stokes) relativistic hydrodynamics and discuss the role of the equation of state (Section~\ref{sec:bjorken_ns}), before moving on to 0+1D Bjorken second-order (Israel-Stewart-type) relativistic hydrodynamics in Section~\ref{sec:is_ns}.\footnote{
 All figures in Sections~\ref{sec:bjorken_ns} and \ref{sec:is_ns} can be reproduced with codes available online~\cite{plotting_scripts}, which solve numerically the 0+1D hydrodynamic equations. Only minimal modifications are necessary to make similar figures with different choices of $\zeta/s(T)$, $\eta/s(T)$, $\tau_0$, $T_0$, \ldots} We then explore 1+1D Navier-Stokes hydrodynamics in Section~\ref{sec:ns_1_plus_1}. Implications of this work for the study of heavy ion collisions are briefly discussed in Section~\ref{sec:hic}.

\section{0+1D (boost-invariant) fluid in first-order (Navier-Stokes) relativistic hydrodynamics}

\label{sec:bjorken_ns}

We use $\tau$-x-y-$\eta_s$, coordinates, with $\tau=\sqrt{t^2-z^2}$ and $\tanh \eta_s=z/t$, and use the metric convention $g^{\mu\nu}=diag(1,-1,-1,-1)$.

In a system with Bjorken symmetries, the flow velocity is $u^\mu=(u^\tau,u^x,u^y,u^{\eta_s})=(1,0,0,0)$ and the sole hydrodynamic equation\footnote{A summary of the relativistic Navier-Stokes equations can be found in Ref.~\cite[Section 3.4]{Hirano:2008hy} for example.}  is that for energy density $\epsilon(\tau)$:
\begin{equation}
\partial_\tau \epsilon=-\frac{(\epsilon+P)}{\tau} \left[ 1-\frac{4}{3\tau T} \frac{\eta}{s}-\frac{1}{\tau T} \frac{\zeta}{s} \right] .
\end{equation}
with $\eta$ and $\zeta$ the shear and bulk viscosities, $s$ the entropy density, $P$ the pressure, $T$ the temperature, and $\tau$ the time-like coordinate defined above.

For what follows, it is more convenient to write the hydrodynamic equation in terms of $T$, the temperature\footnote{We assume that there is no other conserved quantities; in particular, we assume baryon chemical potential $\mu_B=0$.}:
\begin{equation}
\partial_\tau \ln T(\tau)=-\frac{c_s^2(T)}{\tau} \left[ 1-\frac{V(T)}{\tau T}  \right]
\label{eq:NS_T_conserv_Bj}
\end{equation}
with $c_s$ the speed of sound, and
where we defined the combined viscosity $V(T)$ as
\begin{equation}
V(T)\equiv\left( \frac{4}{3} \frac{\eta}{s}(T)+ \frac{\zeta}{s}(T) \right).
\label{eq:combined_visc_Bjorken}
\end{equation}

In this one-dimensional system with Bjorken symmetries, both shear and bulk viscosities respond to the same space-time gradient, $1/\tau$, making them indistinguishable in the Navier-Stokes limit. This statement is of course independent of the nature (equation of state) of the system.

The solution to Eq.~\ref{eq:NS_T_conserv_Bj} can be written
\begin{eqnarray}
\ln \left( \frac{T(\tau)}{T(\tau_0)} \right) &=&-\int_{\tau_0}^{\tau} d\tau^\prime \frac{c_s^2(T(\tau^\prime))}{\tau^\prime} \nn \\
& & +\int_{\tau_0}^{\tau} d\tau^\prime \frac{c_s^2(T(\tau^\prime))}{\tau^\prime} \frac{V(T(\tau^\prime))}{\tau^\prime T(\tau^\prime)} \; .
\label{eq:NS_T_conserv_Bj_int}
\end{eqnarray}
where $T(\tau)$ is the temperature at time $\tau$. The initial conditions are provided by specifying the temperature $T(\tau_0)$ of the system at time $\tau_0$.

To fix ideas, we begin by discussing the simpler case of a constant speed of sound (e.g. a conformal equation of state). The case of QCD is discussed next.

\subsection{Constant speed of sound}

\label{sec:ns_bjorken_conf}

We can see that viscosity affects the right-hand side of Eq.~\ref{eq:NS_T_conserv_Bj_int} in two ways. Since viscosity modifies the temperature profile $T(\tau)$, the temperature dependence of the speed of sound is probed differently in $\int d\tau^\prime c_s^2(T(\tau^\prime))/\tau^\prime$ than in $\int d\tau^\prime c_s^2(T_I(\tau^\prime))/\tau^\prime$, with $T$ and $T_I$ being the viscous and ideal temperature profile respectively. The effect of viscosity on this term depends directly on how much the system's speed of sound varies as a function of temperature. In the case of a constant speed of sound, this first term has no dependence on viscosity.

In the second term, an effective viscosity $V_{\textrm{eff}}$ can be defined exactly through the mean value theorem:
\begin{equation}
\int_{\tau_0}^{\tau} d\tau^\prime \frac{c_s^2}{\tau^\prime} \frac{V(T(\tau^\prime))}{\tau^\prime T(\tau^\prime)}
=
V_{\textrm{eff}} \int_{\tau_0}^{\tau} d\tau^\prime \frac{c_s^2}{\tau^\prime} \frac{1}{\tau^\prime T(\tau^\prime)}
\label{eq:Veff_bjorken_conformal_tau_mean_value_theorem}
\end{equation}
with $V_{\textrm{eff}}=V(T(\tau^{*}));\; \tau_0\le\tau^*\le\tau$. Recall the definition of the combined viscosity $V(T)$ in Eq.~\ref{eq:combined_visc_Bjorken}. If the (viscous) temperature profile $T(\tau)$ is known, from a numerical solution for example, the exact value of the effective viscosity $V_{\textrm{eff}}$ can be calculated:
\begin{equation}
V_{\textrm{eff}} =\frac{\int_{\tau_0}^{\tau}  \frac{d\tau^\prime}{\tau^\prime} \frac{V(T(\tau^\prime))}{\tau^\prime T(\tau^\prime)}}{\int_{\tau_0}^{\tau}  \frac{d\tau^\prime}{\tau^\prime} \frac{1}{\tau^\prime T(\tau^\prime)}} \; .
\label{eq:Veff_bjorken_conformal_tau}
\end{equation}

Because of the symmetries of the system, there is a one-to-one mapping between the time $\tau$ and the temperature $T$. This mapping does not provide by itself a simpler expression for $V_{\textrm{eff}}$. To simplify Eq.~\ref{eq:Veff_bjorken_conformal_tau}, we use the fact that the effect of viscosity on the temperature profile is generally modest. In this ideal case, the mapping between temperature and $\tau$ is given by $d \ln(\tau)= -c_s^{-2} d \ln(T_I)$ (Eq.~\ref{eq:NS_T_conserv_Bj} with $V=0$), or $\ln(\tau/\tau_0)=-c_s^{-2} \ln(T_I/T_0)$, yielding
\begin{equation}
V_{\textrm{eff}} \approx \frac{\int_{T_I(\tau)}^{T_0} d T^\prime  (T^{\prime})^{\left( c_s^{-2}-2 \right)}  V(T^\prime)}{\int_{T_I(\tau)}^{T_0} d T^\prime (T^{\prime})^{\left( c_s^{-2}-2 \right)}} \; .
\label{eq:Veff_bjorken_conformal}
\end{equation}

The more practical Eq.~\ref{eq:Veff_bjorken_conformal} provides a good approximation of the exact Eq.~\ref{eq:Veff_bjorken_conformal_tau} as long as the effective viscosity is modest.

Equation~\ref{eq:Veff_bjorken_conformal} means that the effective viscosity depends on a single moment of the temperature dependence. There exists an infinite family of temperature-dependent $\eta/s(T)$ and $\zeta/s(T)$ whose evolution will be so similar as to make them indistinguishable from each other, and Eq.~\ref{eq:Veff_bjorken_conformal} provides the definition for this family of viscosities. 

\subsection{QCD equation of state}

\label{sec:ns_bjorken_qcd}

\begin{figure}[tb]
	\centering
	\includegraphics[width=0.5\textwidth]{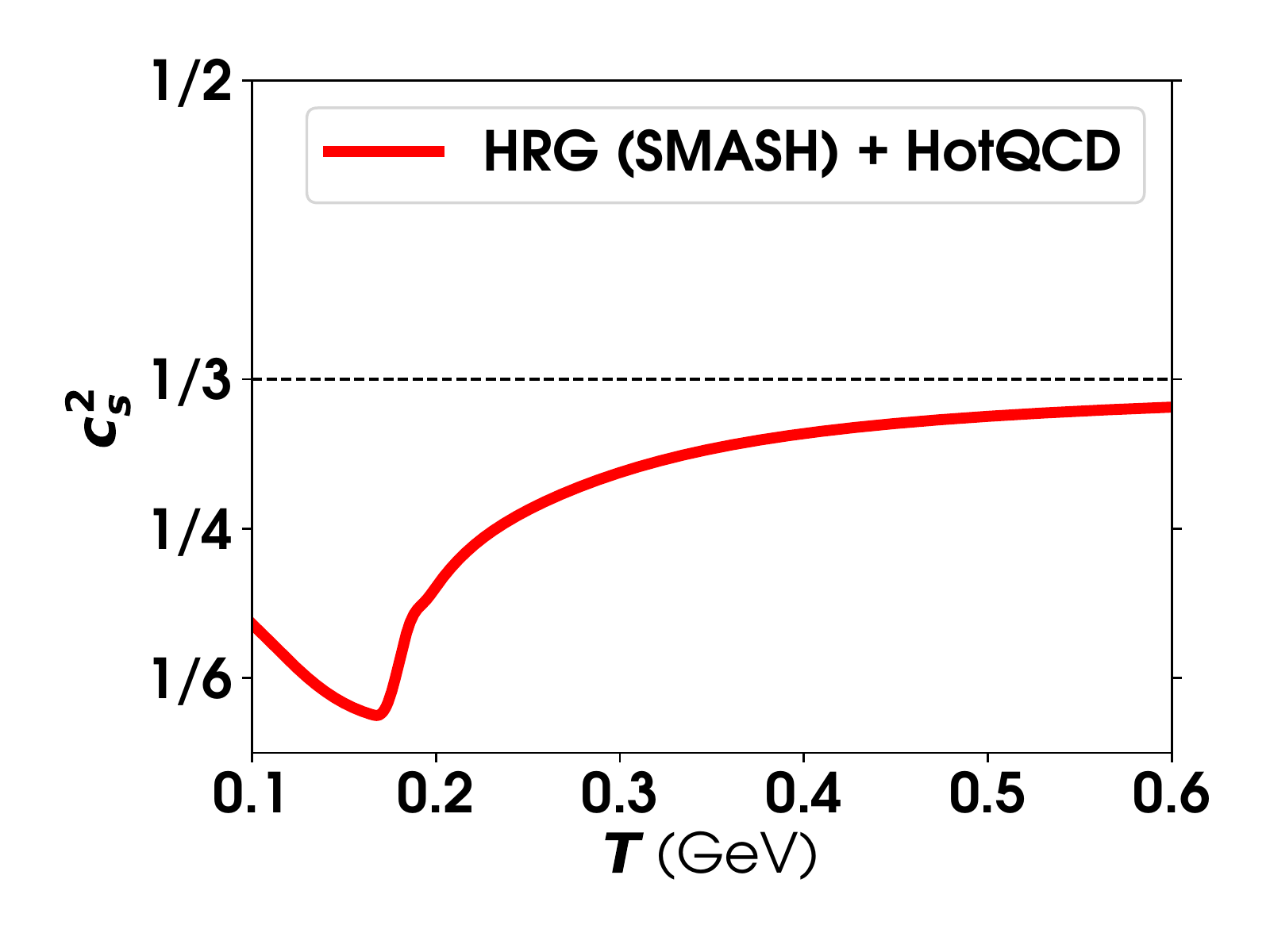}
	\caption{Speed of sound squared as a function of temperature, used in this work. It is based on a recent lattice calculation of the equation of state of QCD~\cite{Borsanyi:2013bia}, matched at low temperature to a hadron resonance gas.}
	\label{fig:cs2}
\end{figure}

A calculation of the speed of sound of QCD\footnote{This speed of sound is obtained by matching a lattice calculation of the equation of state at high temperature~\cite{Borsanyi:2013bia} to a hadron resonance gas at lower temperature (see Refs.~\cite{Bernhard:2018hnz,eos_code} for details on the matching; the particle content of the hadron resonance gas is consistent with that of the SMASH hadronic transport~\cite{Weil:2016zrk}). No uncertainties are shown, as they are not relevant for this work. An uncertainty band can be found in Ref.~\cite{Borsanyi:2013bia}, for example.} is shown in Fig.~\ref{fig:cs2}. The speed of sounds varies slowly above a temperature of $300$~MeV, but changes more rapidly when temperatures reach $\sim 200$~MeV; a minimum is reached around $180$~MeV.

Referring back to Eq.~\ref{eq:NS_T_conserv_Bj_int}, as discussed above, we see that a temperature-dependent speed of sound introduces two major differences. As in the constant $c_s$ scenario, viscosity enters directly in the second term of Eq.~\ref{eq:NS_T_conserv_Bj_int}. With a temperature-dependent speed of sound, viscosity also affects the first term of Eq.~\ref{eq:NS_T_conserv_Bj_int}, through $c_s^2(T(\tau))$. This makes an exact definition of effective viscosity more challenging. Nevertheless it is reasonable to expect that the dominant effect of viscosity is from the second term of Eq.~\ref{eq:NS_T_conserv_Bj_int}, given that it depends directly on the viscosity $V(T)$. This can be seen more clearly if we consider an iterative solution to Eq.~\ref{eq:NS_T_conserv_Bj_int}. In absence of viscosity, Eq.~\ref{eq:NS_T_conserv_Bj_int} reduces to:
\begin{eqnarray}
\ln \left( \frac{T_I(\tau)}{T(\tau_0)} \right) &=&-\int_{\tau_0}^{\tau} d\tau^\prime \frac{c_s^2(T_I(\tau^\prime))}{\tau^\prime} \;.
\label{eq:NS_T_conserv_Bj_int_ideal}
\end{eqnarray}

Inserting this ideal solution, $T_I(\tau)$, on the right-hand side of Eq.~\ref{eq:NS_T_conserv_Bj_int} provides a first iterative solution for the effect of viscosity on the temperature profile:
\begin{eqnarray}
\ln \left( \frac{T(\tau)}{T(\tau_0)} \right) &=&-\int_{\tau_0}^{\tau} d\tau^\prime \frac{c_s^2(T_I(\tau^\prime))}{\tau^\prime} \nn \\
& & +\int_{\tau_0}^{\tau} d\tau^\prime \frac{c_s^2(T_I(\tau^\prime))}{\tau^\prime} \frac{V(T_I(\tau^\prime))}{\tau^\prime T_I(\tau^\prime)} \; .
\label{eq:NS_T_conserv_Bj_int_first_iteration}
\end{eqnarray}

In this first iterative solution, the effect of viscosity enters only in the second term. At this level of approximation, the same steps used in the previous section can be followed, from the application of the mean value theorem in Eq.~\ref{eq:Veff_bjorken_conformal_tau_mean_value_theorem} to the approximate ideal $\tau-T$ mapping in Eq.~\ref{eq:Veff_bjorken_conformal}. The principal difference with the constant-speed-of-sound case is that Eqs.~\ref{eq:Veff_bjorken_conformal_tau_mean_value_theorem} and \ref{eq:Veff_bjorken_conformal_tau} are exact with $c_s$ constant, while they are already approximate with a temperature-dependent $c_s(T)$.

We begin with
\begin{equation}
V_{\textrm{eff}} \approx \frac{\int_{\tau_0}^{\tau} d\tau^\prime \frac{c_s^2(T_I(\tau^\prime))}{\tau^\prime} \frac{V(T_I(\tau^\prime))}{\tau^\prime T_I(\tau^\prime)}}{ \int_{\tau_0}^{\tau} d\tau^\prime \frac{c_s^2(T_I(\tau^\prime))}{\tau^\prime} \frac{1}{\tau^\prime T_I(\tau^\prime)}} \; .
\label{eq:Veff_bjorken_qcd_tau_mean_value_theorem}
\end{equation}

Equation~\ref{eq:NS_T_conserv_Bj_int_ideal} can be rewritten as
\begin{equation}
\tau_I(T) = \tau_0 \exp \left[ - \int_{T_0}^{T} \frac{d T^\prime}{T^\prime} c_s^{-2}(T^\prime) \right] \; .
\end{equation}

Because the speed of sound is a relatively slowly varying function and because temperature follows approximately a power law, we use a logarithmic trapezoid rule to obtain
\begin{equation}
\tau_I(T) \approx \tau_0 \left(\frac{T_0}{T}\right)^{c_s^{-2}\left(\sqrt{T_0 T}\right)}
\label{eq:tau_vs_T_ideal}
\end{equation}
which, combined with the steps we used to obtain Eq.~\ref{eq:Veff_bjorken_conformal}, yields
\begin{equation}
V_{\textrm{eff}} \approx \frac{\int_{T(\tau)}^{T_0}  d T^\prime \left(\frac{T^\prime}{T_0}\right)^{c_s^{-2}\left(\sqrt{T_0 T^\prime}\right)-2} V(T^\prime) }{ \int_{T(\tau)}^{T_0}  d T^\prime \left(\frac{T^\prime}{T_0}\right)^{c_s^{-2}\left(\sqrt{T_0 T^\prime}\right)-2}  } \; .
\label{eq:Veff_bjorken_qcd}
\end{equation}

It is worth emphasizing that even in this Bjorken hydrodynamics with strong imposed symmetries, a series of approximations are necessary to obtain a simple definition of effective viscosity, highlighting the challenge of a general definition of the concept.

With a non-trivial speed of sound, the effective viscosity from Eq.~\ref{eq:Veff_bjorken_qcd} is no longer a simple moment of $V(T)$. There is still a family of temperature-dependent viscosities $V(T)$ that have the same numerical viscosity, but these must be evaluated numerically because of the non-triviality of the speed of sound.

We emphasize again that shear and bulk viscosities are indistinguishable in Bjorken Navier-Stokes hydrodynamics. In practice, to better understand Eq.~\ref{eq:Veff_bjorken_qcd} and study it numerically, we look at shear and bulk viscosities separately.

\paragraph{Shear viscosity}

\label{sec:ns_bjorken_qcd_shear}

\begin{figure}[tb]
	\centering
	\includegraphics[width=0.5\textwidth]{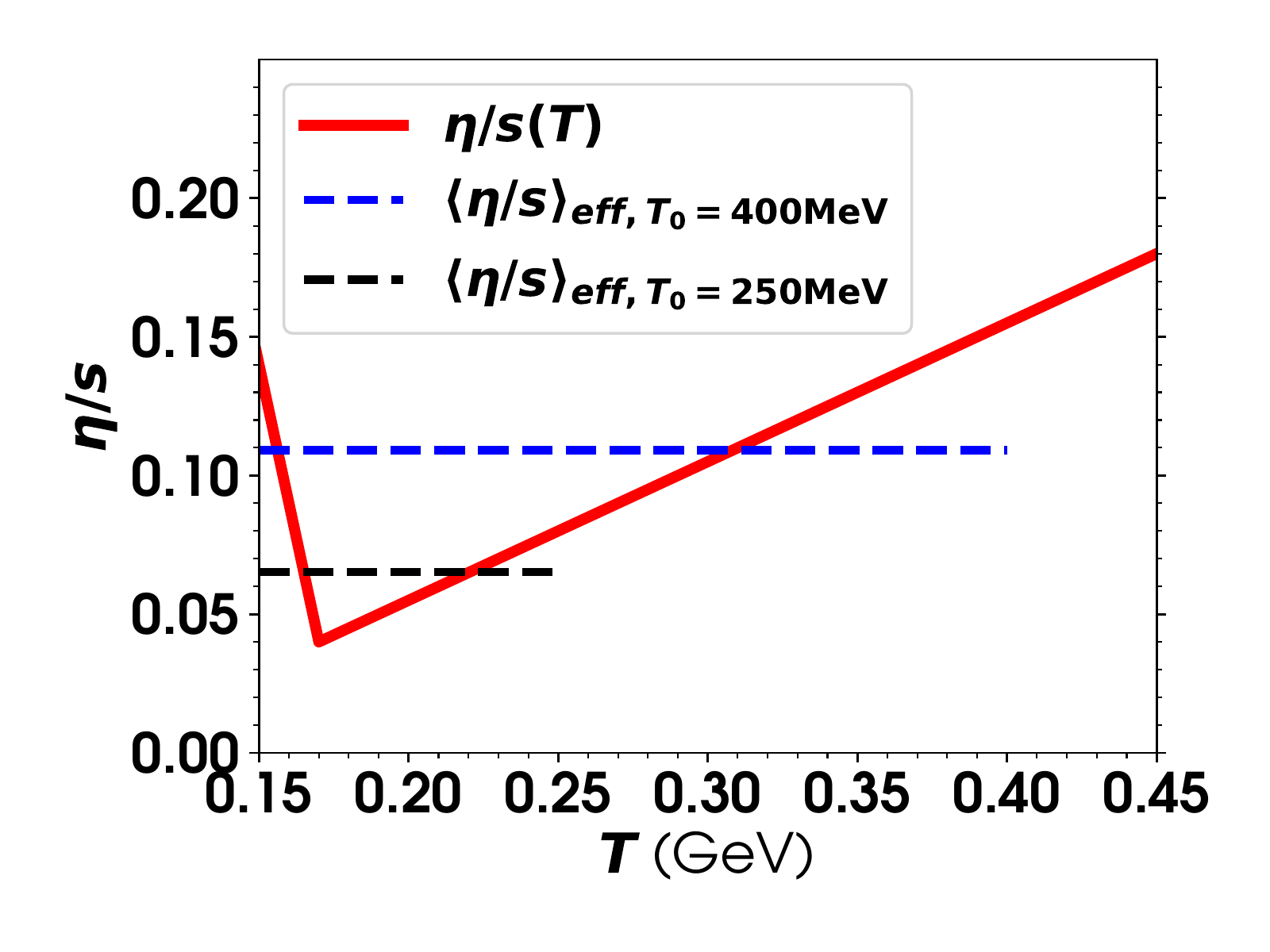}
	\caption{
		Example of temperature-dependent $\eta/s$ with effective shear viscosities corresponding to Bjorken evolutions with $T_0=250$ and $400$ MeV with the QCD equation of state (Fig.~\ref{fig:cs2}). The dashed lines are evaluated with Eq.~\ref{eq:Veff_bjorken_qcd}.
	}
	\label{fig:bjorken_effective_shear_qcd}
\end{figure}

\begin{figure}[tb]
	\centering
	\includegraphics[width=0.5\textwidth]{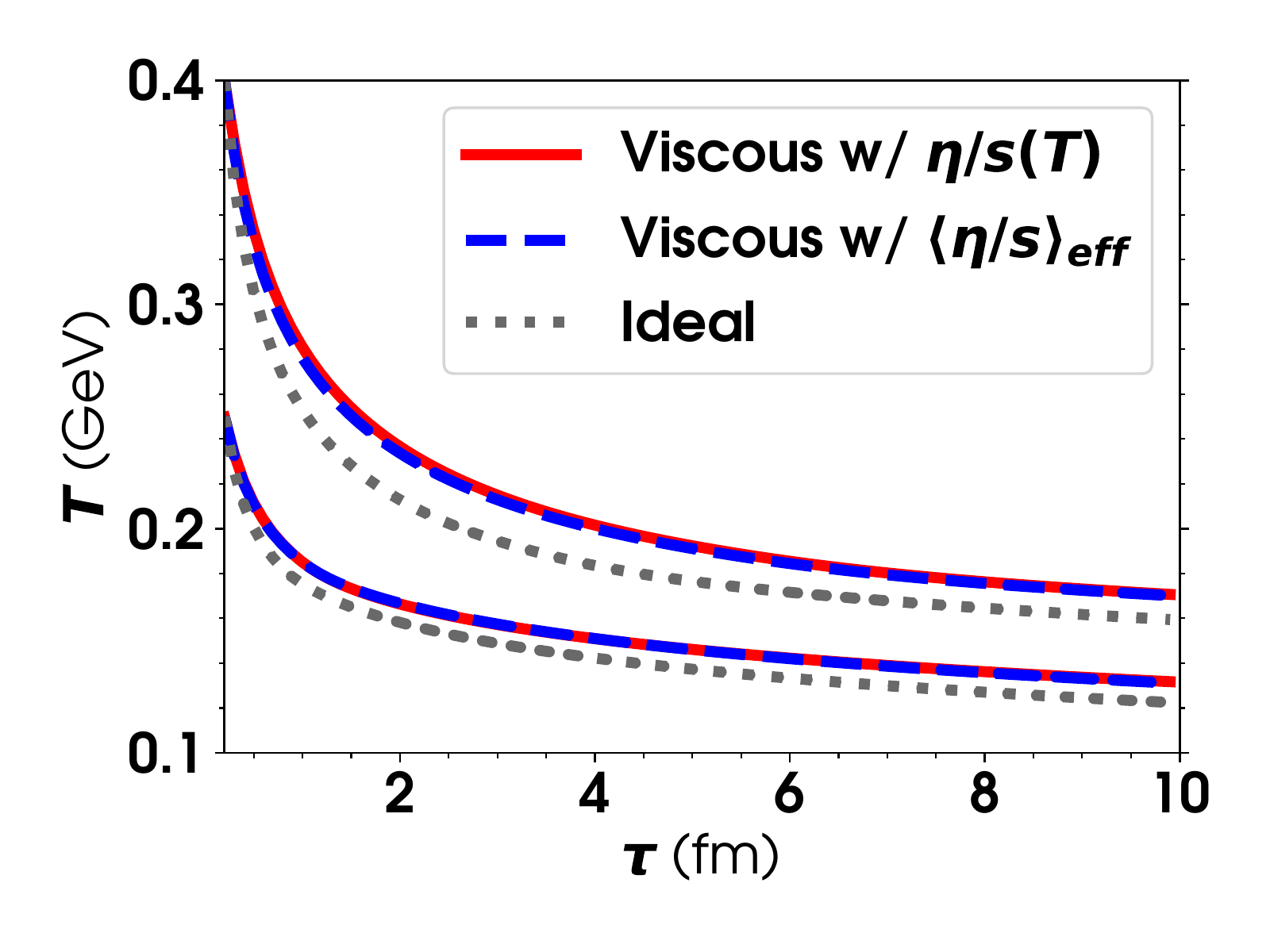}
	\caption{
		Temperature as a function of time $\tau$ for Bjorken hydrodynamics with the effective (Eq.~\ref{eq:Veff_bjorken_qcd}) and temperature-dependent $\eta/s(T)$ shown in Figure~\ref{fig:bjorken_effective_shear_qcd}. Two different initial temperatures $T_0$ are shown, $250$~MeV (lower curves) and $400$~MeV (upper curves). The final temperature is $T_f=150$~MeV in this example, meaning that the effective viscosity is such that this $T_f$ is reached at the same time as for the $\eta/s(T)$ case. The results for ideal Bjorken hydrodynamics is shown for reference.
	}
	\label{fig:bjorken_effective_shear_T_profiles_qcd}
\end{figure}

We choose  a piecewise-linear\footnote{The exact parametrization is:
\begin{equation}
\eta/s(T)=\left\{
\begin{array}{@{}ll@{}}
 0.04 + 5\; (T-T_*), & \textrm{if}\ T<T_* \\
0.04 + 0.5 (T-T_*), & \textrm{if}\ T>T_* 
\end{array}\right.
\end{equation} 
with $T_*=170$~MeV.
} $\eta/s(T)$ with a minimum at $170$~MeV, plotted as a solid line in Fig~\ref{fig:bjorken_effective_shear_qcd} --- a plausible temperature dependence for the shear viscosity to entropy density ratio of QCD. The result\footnote{Note that  Eqs.~\ref{eq:Veff_bjorken_conformal_tau}, \ref{eq:Veff_bjorken_conformal}, \ref{eq:Veff_bjorken_qcd_tau_mean_value_theorem} and \ref{eq:Veff_bjorken_qcd} are equations for the ``combined viscosity'' defined in Eq.~\ref{eq:combined_visc_Bjorken}. Recall the factor $3/4$ when calculating the effective $\eta/s$.} 
of Eq.~\ref{eq:Veff_bjorken_qcd} is shown as a dashed line. The effective viscosity depends on the initial and final temperature of the evolution. In the present case, we choose two values of the initial temperature, $T_0=400$~MeV and $250$~MeV, and we fix the final temperature to $T_f=150$~MeV. The initial time is $\tau_0=0.2$~fm. The result for the effective shear viscosity calculated from Eq.~\ref{eq:Veff_bjorken_qcd} is $\langle \eta/s \rangle_{\textrm{eff}}=0.065$ for $T_0=250$~MeV and $\langle \eta/s \rangle_{\textrm{eff}}=0.11$ for $T_0=400$~MeV. Note that for such a smooth temperature dependence of $\eta/s(T)$, the variation of the speed of sound is a small effect, and $\langle \eta/s \rangle_{\textrm{eff}}$ changes by less than $5$\% in this example if $c_s^{-2}=3$ is used instead of the real speed of sound of QCD.

The temperature evolution obtained by using the effective viscosities from Eq.~\ref{eq:Veff_bjorken_qcd} is shown in Fig.~\ref{fig:bjorken_effective_shear_T_profiles_qcd}. Recall that the effective viscosity is defined such as to obtain the correct temperature at some fixed final time $\tau_f$ (or final temperature $T_f$). In principle, it does not insure that the entire evolution is the same as that obtained with $\eta/s(T)$. In practice, the dependence of the effective viscosity on the final time $\tau_f$ (or temperature $T_f$) is generally small. One way to understand why is to recall the factor of $\left(T^\prime/T_0\right)$ in Eq.~\ref{eq:Veff_bjorken_qcd}, whose exponent $[c_s^{-2}\left(\sqrt{T_0 T^\prime}\right)-2]$ is always larger than $1$: the effective viscosity is generally dominated by contributions at large values of temperature\footnote{In this 0+1D Bjorken symmetric system, high temperatures mean early times. Viscosities respond to spacetime gradients, and time gradients go as $1/\tau$ in a Bjorken evolution: the dominant effect of viscosity is expected to be at early time, or high temperatures.}. This is characteristic of the 0+1D Bjorken fluid hydrodynamics expansion assumed in this section. Fluid expanding in the transverse directions can be expected to have a larger dependence on this final time $\tau_f$ (or temperature $T_f$).

\paragraph{Bulk viscosity}

\label{sec:ns_bjorken_qcd_bulk}

\begin{figure}[tb]
	\centering
	\includegraphics[width=0.5\textwidth]{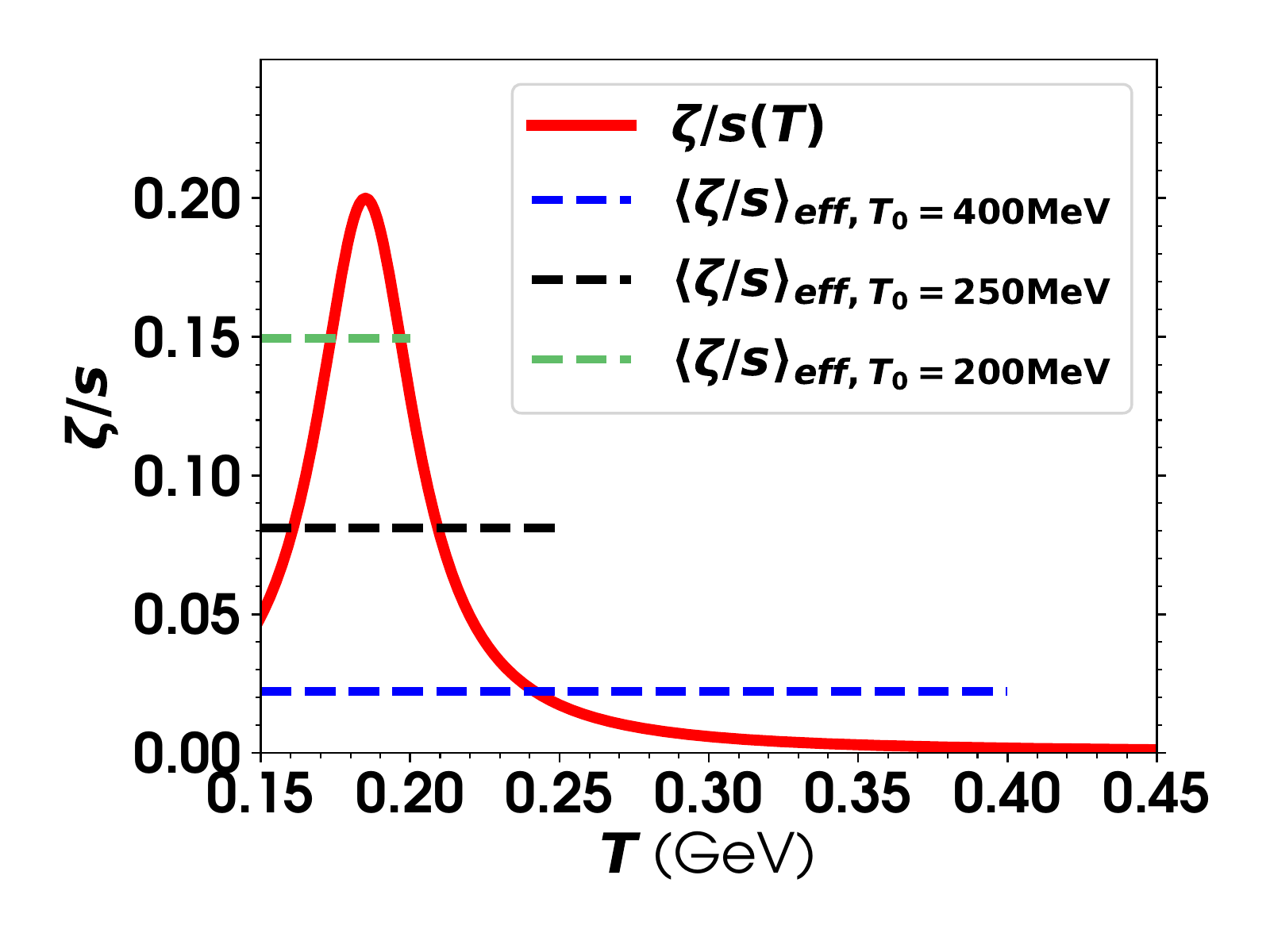}
	\caption{
		Example of temperature-dependent $\zeta/s$ with effective bulk viscosities corresponding to Bjorken evolutions with $T_0=200$, $250$ and $400$ MeV with the QCD equation of state (Fig.~\ref{fig:cs2}). The dashed lines are evaluated with Eq.~\ref{eq:Veff_bjorken_qcd}.
	}
	\label{fig:bjorken_effective_bulk_qcd}
\end{figure}

At high temperature, the bulk viscosity of QCD is much smaller than its shear viscosity~\cite{Arnold:2006fz}, a consequence of QCD being nearly conformal in this limit. There have been discussions that bulk viscosity is not necessarily small in the deconfinement region, $T\sim 150-250$~MeV, where shear viscosity may reach a minimum while bulk viscosity takes a larger value, possibly a narrow peak.\footnote{See Ref.~\cite{Noronha-Hostler:2015qmd} for a summary and a visualization of multiple recent calculations of shear and bulk viscosities.}

As an example, we use\footnote{\label{footnote:bulk_param}
\begin{equation}
\zeta/s(T)=\frac{\langle \zeta/s \rangle_{max}}{1+(T-T_*)^2/\sigma^2}
\end{equation}
with $\langle \zeta/s \rangle_{max}=0.2$, $T_*=185$~MeV and $\sigma=20$~MeV.
} a Cauchy distribution for $\zeta/s$ with a maximum of $0.2$ at $T=185$~MeV. This parametrization of $\zeta/s(T)$ is shown in Fig.~\ref{fig:bjorken_effective_bulk_qcd} with the effective viscosities computed from Eq.~\ref{eq:Veff_bjorken_qcd}  shown as dashed lines. Three values of the initial temperature are shown, $T_0=200$, $250$ and $400$~MeV, and the initial time and final temperature are the same as in the previous section, $\tau_0=0.2$~fm and $T_f=150$~MeV. The effective viscosities are
\begin{itemize}
	\item $\zetaeff=0.02$ for $T_0=400$~MeV,
	\item $\zetaeff=0.08$ for $T_0=250$~MeV,
	\item $\zetaeff=0.15$ for $T_0=200$~MeV.
\end{itemize}
A smaller effective viscosity can be understood as a viscosity that is more difficult to probe. In a Bjorken expanding fluid, the expansion rate $\theta=1/\tau$ becomes smaller as time increases and temperature decreases: the best way to probe $\zeta/s(T)$ is for the initial temperature to be close to that of the peak of $\zeta/s(T)$, as shown in Fig.~\ref{fig:bjorken_effective_bulk_qcd}.

In fact, for $\zeta/s(T)$ that falls off rapidly at high temperature, one can derive an  approximate relation for effective bulk viscosities with different initial temperature $T_0$. Suppose that $T^{a}_{0}$ and $T^{b}_{0}$ are two different initial temperatures at high enough temperature where $\zeta/s(T)$ does not have significant support. Assuming a conformal equation of state, we see that the numerator of Eq.~\ref{eq:Veff_bjorken_conformal} will be approximately the same for $T^{a}_{0}$ and $T^{b}_{0}$. The difference between the effective bulk viscosities will thus be
\begin{equation}
\frac{\langle \zeta/s \rangle_{\textrm{eff, a}}}{\langle \zeta/s \rangle_{\textrm{eff, b}}} \approx \frac{\left(T_0^b\right)^2-\left(T_f\right)^2}{\left(T_0^a\right)^2-\left(T_f\right)^2}.
\label{eq:zeta_eff_ratio}
\end{equation}
Looking back at Fig.~\ref{fig:bjorken_effective_bulk_qcd}, taking $T^{a}_{0}=250$~MeV and $T^{b}_{0}=400$~MeV, Eq.~\ref{eq:zeta_eff_ratio} predicts $3.4$ as ratio of the effective bulk viscosity, while the actual ratio is $3.7$.

\begin{figure}[tb]
	\centering
	\includegraphics[width=0.5\textwidth]{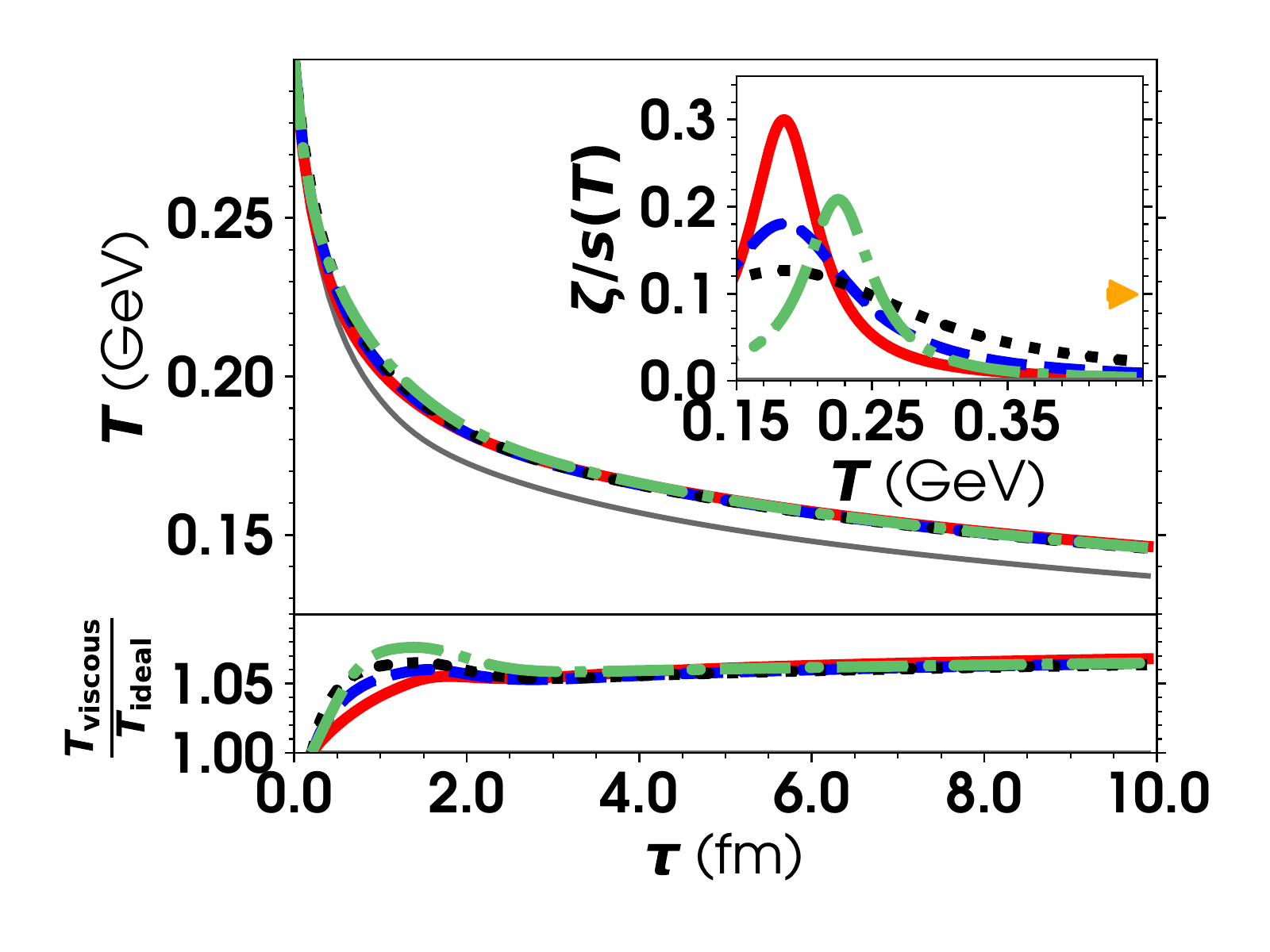}
	\caption{
		Example of different temperature-dependent $\zeta/s$ with equivalent effective bulk viscosities, which lead to similar temperature evolutions. The parameters for the Bjorken hydrodynamics are $T_0=300$~MeV, $\tau_0=0.2$~fm and $T_f=150$~MeV. The value of the effective $\zeta/s$ is indicated by the arrow on the right axis of the inset. The ideal result is shown in grey for reference.
	}
	\label{fig:bjorken_effective_bulk_qcd_equiv}
\end{figure}

In Fig.~\ref{fig:bjorken_effective_bulk_qcd_equiv}, we show a family of temperature-dependent $\zeta/s$ with equivalent effective bulk viscosities, as evaluated with Eq.~\ref{eq:Veff_bjorken_qcd}. While the temperature evolution is not strictly identical (given that Eq.~\ref{eq:Veff_bjorken_qcd} is not an exact definition of effective viscosity), it would be difficult to tell these different temperature dependences of $\zeta/s(T)$ apart from their temperature profiles.
Note that all these equivalent parametrizations of $\zeta/s(T)$ were obtained assuming the same functional form for the temperature dependence (shown in footnote~\ref{footnote:bulk_param}), which is a practical choice but not a necessary one.

\subsection{Alternative approach: effective viscosity as optimization of a hypersurface}

As emphasized in the sections above, our definition of viscosity does not technically require the temperature evolution to be the same over a wide range of time: it is defined such that the temperature profiles are the same at a fixed point in $\tau$ that we call $\tau_f$ (or equivalently at a fixed temperature $T_f$).
As discussed previously, this distinction is in general not relevant, since the effect of viscosity tends to concentrate at earlier times (because spacetime gradients are larger there); this generally leads to similar temperature profiles over large ranges of $\tau$.

It is nevertheless interesting to recast the question of effective viscosity differently. Suppose that we define a constant-temperature hypersurface $\Sigma(T_f,\eta/s(T),\zeta/s(T))$, with $T_f$ the temperature of the fluid on this hypersurface. In 3+1D, $\Sigma$ would be a surface in 4 dimensions; in 0+1D, this hypersurface reduces to a single value of $\tau$: $\tau_f(T_f,\eta/s(T),\zeta/s(T))$.
The effective viscosities $\etaeff$ and $\zetaeff$ can be defined so as to minimize the differences between the hypersurface:
\begin{equation}
min\{\tau_f(T_f,\etaeff,\zetaeff)-\tau_f(T_f,\eta/s(T),\zeta/s(T))\}
\end{equation}

This approach leads to a similar definition of effective viscosity; the full derivation can be found in Appendix~\ref{sec:effective_visc_Bjorken_optimization}. An interesting summary of the derivation is the following approximate relation between the effective viscosity and the evolution time necessary to reach a given final temperature $T_f$:
\begin{eqnarray}
\frac{\tau_{viscous}(T_f)}{\tau_{ideal}(T_f)}&\approx& \exp \left[\frac{3 V_\textrm{eff}(T_f)}{2 \tau_0 T_0} \right] 
\end{eqnarray}
which can be inverted to give
\begin{equation}
V_\textrm{eff}(T_f) \approx \frac{2}{3} \tau_0 T_0 \left( \frac{\tau_{viscous}(T_f)}{\tau_{ideal}(T_f)} -1 \right) \; .
\end{equation}
Note that we used the combined viscosity from Eq.~\ref{eq:combined_visc_Bjorken} to regroup shear and bulk viscosity together.
Since viscosity leads to entropy production, it takes longer in the viscous case $[\tau_{viscous}(T_f)]$ to cool down to $T_f$ than it does in the ideal case $[\tau_{ideal}(T_f)]$. Any temperature-dependent $\eta/s(T)$ or $\zeta/s(T)$ that reaches $T_f$ at the same $\tau_{viscous}(T_f)$ can be said to have the same effective viscosity. This approach could be interesting for the study of heavy ion collisions, where often only the value of the hydrodynamic fields on a constant temperature hypersurface matters.\footnote{Note that while constant temperature hypersurfaces are used almost universally in hydrodynamic studies of heavy ion collisions, hypersurfaces based on other criteria would in theory be better justified. See for example Ref.~\cite{Ahmad:2016ods}.}

\section{0+1D (boost-invariant) fluid in second-order (Israel-Stewart-type) relativistic hydrodynamics}

\label{sec:is_ns}

In the previous section, we saw that there exists families of temperature-dependent $\eta/s(T)$ and $\zeta/s(T)$ that lead to essentially indistinguishable hydrodynamic evolution.
This was for first-order hydrodynamics with a Bjorken-symmetric fluid --- that is a boost-invariant fluid with no transverse dynamics.
The strong symmetries imposed on the system are one reason for this degeneracy. It is also a consequence of the form of the first-order Navier-Stokes hydrodynamics equations, in which the effect of viscosity only enters through local spacetime gradients.

The situation is different for second-order Israel-Stewart-type hydrodynamics. In this latter case, the shear tensor and bulk pressure follow  relaxation-type equations of motion. The effect of viscosity is still through local spacetime gradients; however the evolution of the shear tensor and bulk pressure depend on their initial value as well as on the relaxation time. For example,  for a fluid with Bjorken symmetries, the bulk pressure $\Pi$ evolves with the equation of motion\footnote{In this example, we assume that there is only bulk viscosity, and that the only additional second-order transport coefficients are $\tau_{\Pi}$ and $\delta_{\Pi \Pi }$. In practice, there could be additional high-order terms, such as $\Pi^2$, terms involving vorticity, \ldots The presence of shear viscosity would also introduce a number of shear-bulk couplings. See Ref.~\cite{Denicol:2012cn} for example. These features are set aside for the proof of principle discussed in this section.}
\begin{equation}
	\tau_{\Pi}  \partial_\tau \Pi=-(\Pi-\Pi_{NS}) -\frac{\delta_{\Pi \Pi }\Pi}{\tau}
\end{equation}
with $\Pi_{NS}=-\zeta \theta=-\zeta/\tau$ is the Navier-Stokes bulk pressure, $\tau_\Pi$ is the bulk relaxation time, and $\delta_{\Pi \Pi }$ is a second-order transport coefficient.
The coupled temperature equation of motion is
\begin{equation}
\partial_\tau \ln T(\tau)=-\frac{c_s^2(T)}{\tau} \left[ 1+\frac{\Pi}{s T}  \right] \; .
\label{eq:IS_T_conserv_Bj}
\end{equation}

If the bulk relaxation time $\tau_\Pi$ is very long, the bulk pressure will remain for a long time at its initial value $\Pi(\tau_0)$. In the opposite scenario where the bulk relaxation time is very short, the bulk pressure will be close to the result from first-order hydrodynamics. In scenarios where $\tau_\Pi$ and $\Pi(\tau_0)$ take extreme values, the effect of $\zeta/s(T)$ on the temperature evolution might be smaller than the effect of $\Pi(\tau_0)$.

In terms of $\hat{\Pi}=\Pi/(s T)$, the equation of motion for the bulk pressure can be written
\begin{equation}
\partial_\tau \hat{\Pi}=-\frac{(\hat{\Pi}-\hat{\Pi}_{NS})}{\tau_\Pi} +\frac{\hat{\Pi}}{\tau} \left[ 1 + c_s^{2} -\frac{\delta_{\Pi\Pi}}{\tau_{\Pi}} \right]+\frac{\hat{\Pi}^2}{\tau} (1+c_s^{2})
\label{eq:IS_T_conserv_Bj_piHat}
\; .
\end{equation}
We used Eq.~\ref{eq:IS_T_conserv_Bj} to obtain Eq.~\ref{eq:IS_T_conserv_Bj_piHat}, which is the origin of the quadratic term in $\hat{\Pi}$.
In what follows, we fix $\delta_{\Pi \Pi }$ to the 14-moment approximation result from Ref.~\cite{Denicol:2014vaa}: $\delta_{\Pi \Pi }=(2/3) \tau_\Pi$.

An approximate solution for $\hat{\Pi}$ can be found by neglecting the $\hat{\Pi}^2$ term, approximating the speed of sound as constant\footnote{Choosing $\bar{c}_s^2=1/3$ can appear to be a good option, but it is rarely the best one. As seen in Fig.~\ref{fig:cs2}, the speed of sound of QCD is always smaller than $1/3$; it is better to expand around smaller value, like $1/4$ or $1/5$, than to expand around the asymptotic $1/3$ value. A simple example of this is given in Appendix~\ref{sec:appendix_bjorken} in the context of finding an approximate solution for the temperature in Bjorken hydrodynamics with the QCD equation of state.} ($\bar{c}_s^2$), and assuming that $\tau_\Pi$ is a constant. The resulting expression for $\hat{\Pi}$ is an integral whose dominant contribution can be obtained by integration by part. Under these approximations, the result takes the simple form:
\begin{equation}
\hat{\Pi}(\tau) \approx \left [ \hat{\Pi}(\tau_0) - \hat{\Pi}_{NS}(\tau_0) \right] \left( \frac{\tau}{\tau_0}\right)^{\frac{1}{3}+\bar{c}_s^2} e^{-\frac{\tau-\tau_0}{\tau_\Pi}}  +  \hat{\Pi}_{NS}(\tau) \; .
\label{eq:Pi_hat_approx}
\end{equation}

Equation~\ref{eq:Pi_hat_approx} highlights some of the differences between first and second order hydrodynamics. As expected, it shows that the relaxation toward Navier-Stokes is faster if the bulk pressure is initialized at its Navier-Stokes value. Evidently Equation~\ref{eq:Pi_hat_approx} is approximate, and having $\hat{\Pi}(\tau_0)=\hat{\Pi}_{NS}(\tau_0)$ does not actually imply instantaneous relaxation to Navier-Stokes for the full solution. 
While approximate, Equation~\ref{eq:Pi_hat_approx} can be used to gain intuition on the relation between $\hat{\Pi}(\tau_0)$, $\tau_\Pi$ and the apparent viscosity of a fluid.

As in the Navier-Stokes case (Eqs.~\ref{eq:NS_T_conserv_Bj_int_first_iteration} and \ref{eq:Veff_bjorken_qcd_tau_mean_value_theorem}), the mean value theorem can be used to define an approximate effective viscosity. The most direct definition might be
\begin{equation}
\int_{\tau_0}^{\tau} d\tau^\prime \frac{c_s^2(T(\tau^\prime))}{\tau^\prime} \hat{\Pi}(\tau^\prime) = \langle \hat{\Pi} \rangle_{\textrm{eff}} \int_{\tau_0}^{\tau} d\tau^\prime \frac{c_s^2(T(\tau^\prime))}{\tau^\prime} \; .
\end{equation}
To connect with the Navier-Stokes case more easily, we use an alternative definition\footnote{The Navier-Stokes bulk pressure is a negative quantity in our expanding $0+1$D fluid: $\Pi_{NS}=-(\zeta/s)/(T \tau)$. The additional minus signs in Eq.~\ref{eq:pre_Veff_IS_bjorken_qcd_tau_mean_value_theorem} makes it match the definition of effective viscosity from Section~\ref{sec:bjorken_ns}.}:
\begin{equation}
- \int_{\tau_0}^{\tau} d\tau^\prime \frac{c_s^2(T(\tau^\prime))}{\tau^\prime} \hat{\Pi}(\tau^\prime) = \langle - \tau T \hat{\Pi} \rangle_{\textrm{eff}} \int_{\tau_0}^{\tau} d\tau^\prime \frac{c_s^2(T(\tau^\prime))}{\tau^\prime} \frac{1}{\tau^\prime T(\tau^\prime)}
\label{eq:pre_Veff_IS_bjorken_qcd_tau_mean_value_theorem}
\end{equation}
which is strictly equivalent to Eq.~\ref{eq:Veff_bjorken_qcd_tau_mean_value_theorem} in the Navier-Stokes case ($\hat{\Pi}=\hat{\Pi}_{NS}$).

Inserting Eq.~\ref{eq:Pi_hat_approx} in Eq.~\ref{eq:pre_Veff_IS_bjorken_qcd_tau_mean_value_theorem} and following the same steps as in Section~\ref{sec:bjorken_ns}, we find
\begin{equation}
\langle -\tau T \hat{\Pi} \rangle_{\textrm{eff}} \approx \left [ \hat{\Pi}(\tau_0) - \hat{\Pi}_{NS}(\tau_0) \right] Y(T,T_0) + \langle \zeta/s \rangle_{\textrm{eff}}^{\textrm{NS}}
\label{eq:Veff_IS_bjorken_qcd}
\end{equation}
with
\begin{widetext}
\begin{equation}
Y(T,T_0) \approx - \frac{\int_{T}^{T_0} \frac{d T^\prime}{T^\prime} \exp\left[ -\frac{\tau_0}{\tau_\Pi} \left[ \left( \frac{T_0}{T^\prime} \right)^{c_s^{-2}(\sqrt{T_0 T^\prime})}-1 \right] \right] \left( \frac{T_0}{T^\prime} \right)^{\left(\frac{1}{3}+\bar{c}_s^2\right) c_s^{-2}(\sqrt{T_0 T^\prime})}}{ \frac{1}{\tau_0 T_0} \int_{T}^{T_0}  \frac{d T^\prime}{T^\prime} \left(\frac{T^\prime}{T_0}\right)^{c_s^{-2}\left(\sqrt{T_0 T^\prime}\right)-1}   } \; .
\label{eq:Y_factor}
\end{equation}
\end{widetext}

The meaning of Eq.~\ref{eq:Veff_IS_bjorken_qcd} is the following: combinations of $\zeta/s(T)$, $\tau_\Pi$ and $\hat{\Pi}(\tau_0)$ that yields the same value for Eq.~\ref{eq:Veff_IS_bjorken_qcd} will have similar temperature profiles. Note that more approximations (in particular, Eq.~\ref{eq:Pi_hat_approx}) were necessary to obtain this definition of ``effective viscosity'', compared to the Navier-Stokes case; Eq.~\ref{eq:Veff_IS_bjorken_qcd} may not produce temperature profiles as similar as those seen in Section~\ref{sec:bjorken_ns}. With this caution in mind, we proceed with two examples which use Eq.~\ref{eq:Veff_IS_bjorken_qcd} to better understand the relation between $\zeta/s(T)$, $\tau_\Pi$ and $\hat{\Pi}(\tau_0)$.

\paragraph{Breaking the degeneracy of Navier-Stokes effective viscosities}

In Section~\ref{sec:bjorken_ns}, we discussed that different parametrizations of $\zeta/s(T)$ could produce nearly indistinguishable temperature evolutions, for Bjorken Navier-Stokes hydrodynamics, if their effective viscosity $\langle \zeta/s \rangle_{\textrm{eff}}^{\textrm{NS}}$ are the same. In principle, this degeneracy is broken in second-order hydrodynamics.
In  Eq.~\ref{eq:Veff_IS_bjorken_qcd},  there is an additional dependence on $\zeta/s(T_0)$ in the first term, through $\hat{\Pi}_{NS}(\tau_0)=-[\zeta/s(T_0)]/(\tau_0 T_0)$.

In practice, the degeneracy between the different $\zeta/s(T)$ would only be significantly broken  by second-order hydrodynamics in very specific cases. It requires\footnote{If $\left|\hat{\Pi}(\tau_0)\right| \gg \left| \hat{\Pi}_{NS}(\tau_0) \right|$, it is trivial to see that Eq.~\ref{eq:Veff_IS_bjorken_qcd} is essentially independent of $\hat{\Pi}_{NS}(\tau_0)$. } 
\begin{equation}
\left|\hat{\Pi}(\tau_0)\right| \sim \left| \hat{\Pi}_{NS}(\tau_0) \right|,
\end{equation}
as well as the first term of Eq.~\ref{eq:Veff_IS_bjorken_qcd} to be large compared to $\langle \zeta/s \rangle_{\textrm{eff}}^{\textrm{NS}}$.
We verified numerically that in the case of Fig.~\ref{fig:bjorken_effective_bulk_qcd_equiv}, for example, the temperature evolution for the different $\zeta/s(T)$ remains degenerate for most choices of $\hat{\Pi}_{NS}(\tau_0)$ and $\tau_\Pi$.

\paragraph{Mimicking viscosity with out-of-equilibrium initial conditions}

\begin{figure}[tb]
	\centering
	\includegraphics[width=0.5\textwidth]{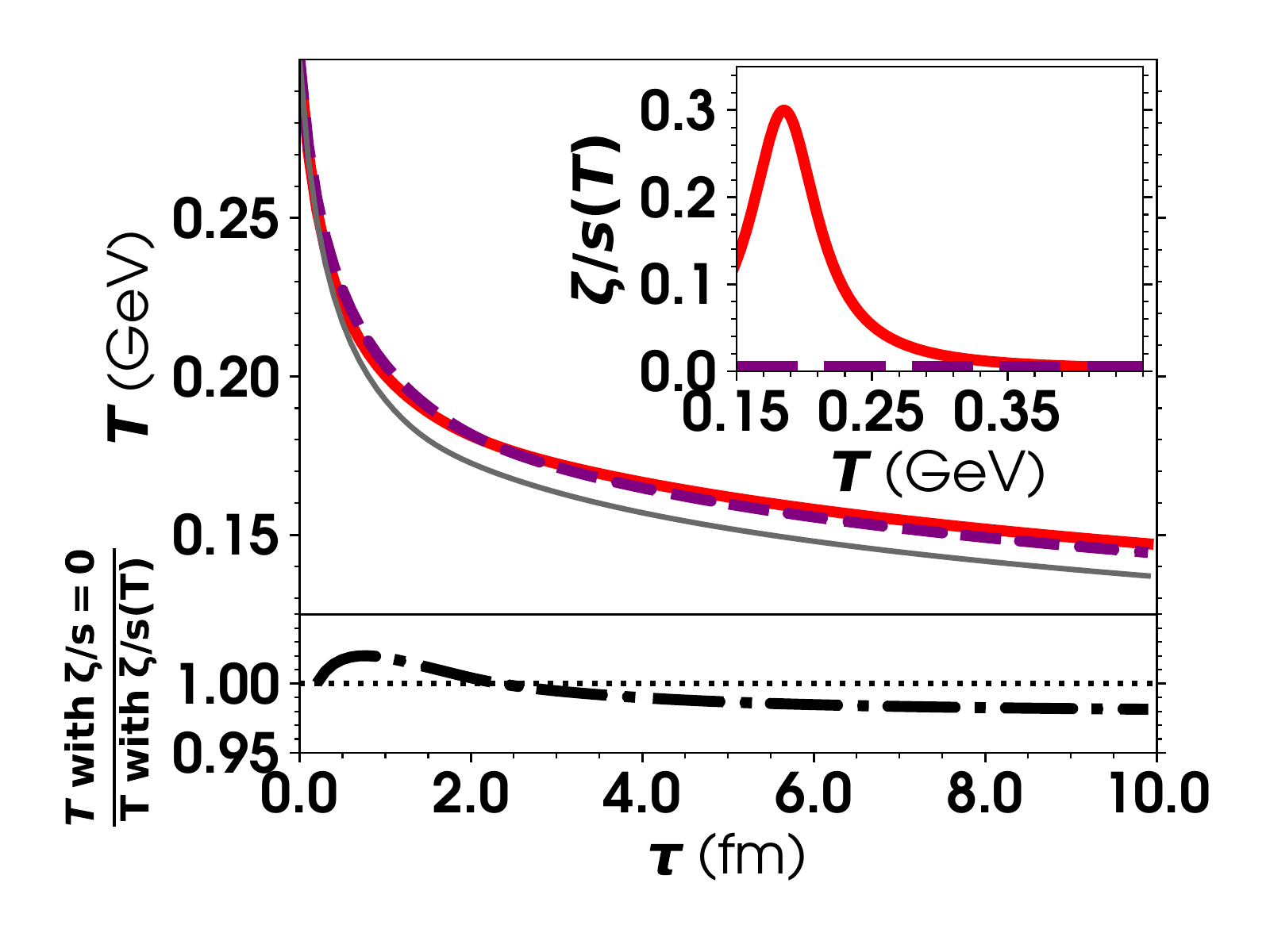}
	\caption{
		Comparisons of temperature profiles obtained with (i) a temperature-dependent $\zeta/s$ and the Navier-Stokes value for the initial bulk pressure, and (ii) an asymptotically small $\zeta/s$ and an initial bulk pressure given by Eq.~\ref{eq:PiHat0_Veff_relaxation}. The parameters for the Bjorken hydrodynamics are $T_0=300$~MeV, $\tau_0=0.2$~fm and $T_f=150$~MeV. The QCD equation of state is used. The relaxation time is constant, $\tau_\Pi=1$~fm. The value of $\bar{c}_s^2$ used in Eq.~\ref{eq:Y_factor} was $1/4$;  we verified that similar results can be obtained with other values of $\bar{c}_s^2$.}
	\label{fig:is_bjorken_effective_bulk_qcd_Pi0_mimic}
\end{figure}

The first term of Eq.~\ref{eq:Veff_IS_bjorken_qcd} effectively quantifies how a viscosity can be mimicked by a non-equilibrium initial value of the bulk pressure. Consider any of the parametrizations of $\zeta/s(T)$ shown in Fig.~\ref{fig:bjorken_effective_bulk_qcd_equiv}. They have an effective viscosity of $\langle \zeta/s \rangle_{\textrm{eff}}^{\textrm{NS}} \approx 0.1$. As discussed above, for most choices of $\hat{\Pi}_{NS}(\tau_0)$ and $\tau_\Pi$, this degeneracy between the different parametrizations of $\zeta/s(T)$ will remain.

Suppose an extreme case where the bulk viscosity of QCD is negligibly small. Even in this scenario, one can obtain a temperature evolution similar to those seen in Fig.~\ref{fig:bjorken_effective_bulk_qcd_equiv} with $\langle \zeta/s \rangle_{\textrm{eff}}^{\textrm{NS}} \approx 0.1$, by using
\begin{equation}
\hat{\Pi}(\tau_0) = \frac{\langle \zeta/s \rangle_{\textrm{eff}}^{\textrm{NS}}}{Y(T,T_0)} \; .
\label{eq:PiHat0_Veff_relaxation}
\end{equation}

If we apply this prescription to the example shown in Section~\ref{sec:ns_bjorken_qcd_bulk} (Fig.~\ref{fig:bjorken_effective_bulk_qcd_equiv}), we obtain Fig.~\ref{fig:is_bjorken_effective_bulk_qcd_Pi0_mimic}: 
if a proper value of $\hat{\Pi}(\tau_0)$ is used (namely Eq.~\ref{eq:PiHat0_Veff_relaxation}),
a temperature profile similar to that obtained with a non-trivial $\zeta/s(T)$ can be obtained with no bulk viscosity whatsoever. The temperature profiles are not identical, reflecting the limitations of Eq.~\ref{eq:Veff_IS_bjorken_qcd}. Nevertheless, this example highlights the partial degeneracy that exists between the out-of-equilibrium initial conditions and the transport coefficients.

\section{Effective viscosities beyond 0+1D}

\label{sec:ns_1_plus_1}

While the previous section assumed a system with perfect Bjorken symmetries (no transverse dynamics), its conclusions are expected to hold in systems with mild transverse gradients, at sufficiently early times. %

We first explore this simpler scenario, before moving on to a more general setting. We limit the whole section's discussion to first-order (Navier-Stokes) hydrodynamics. Derivations are performed for a general speed of sound; all numerical results shown use the QCD equation of state.

\subsection{Effective viscosities in a cylindrically-symmetric Bjorken system: small gradients and early time limit}

As an example, suppose a system whose initial temperature has a Gaussian profile in the transverse plane while still boost-invariant in the longitudinal direction:
$$T(\tau_0,r,\eta_s)=T_0 \exp(-r^2/\sigma^2)$$
at initial time $\tau=\tau_0$, with maximum initial temperature\footnote{With $\sigma$ of order $5-10$~fm and $T_0(\tau_0=0.2\textrm{ fm})\sim 400-600$~MeV, this scenario is closer to that encountered in ultrarelativistic head-on collisions of heavy ions at the RHIC and the LHC. It must be emphasized, however, that heavy ion collisions never have such a high degree of symmetry in the transverse plane at early times.} $T_0$ and width $\sigma$. The transverse direction is $r=\sqrt{x^2+y^2}$ and $\eta_s$ is the spatial rapidity defined in Section~\ref{sec:bjorken_ns}.
Limiting our discussing to bulk viscosity for simplicity, the equation of motion for the temperature in a cylindrically-symmetric boost-invariant system is given by
\begin{equation}
u^{\tau} \partial_{\tau} \ln T + u^r \partial_{r} \ln T = -c_s^2(T) \left[ \theta - \frac{1}{T} \frac{\zeta}{s} \theta^2 
\right]
\label{eq:NS_T_cyl_eom}
\end{equation}
with
\begin{equation}
\theta=\frac{u^\tau}{\tau} + \frac{u^r}{r} + \partial_\tau u^\tau + \partial_r u^r \; .
\label{eq:theta_cyl}
\end{equation}

Compared to the 0+1 case, the single additional dimensionful scale would be the width $\sigma$ for a constant speed of sound; a non-constant speed of sound introduces an additional scale, something like $\partial c_s^2/\partial T$.
At early times, spatial gradients scale like $1/\sigma$ while temporal gradients scale like $1/\tau$. The expansion rate $\theta$ is thus dominated by the $1/\tau$ term as long as $\tau$ is reasonably smaller than $\sigma$. Thus, for $\tau \ll \sigma$, the temperature can be approximated by
\begin{equation}
T(\tau,r)=T_0(\tau,r) \left( \frac{\tau_0}{\tau} \right)^{c_s^2}
\label{eq:tube_solution}
\end{equation}
for a constant speed of sound.
The derivation of effective viscosity defined in Section~\ref{sec:bjorken_ns} can be applied locally in $r$. Moreover it can be generalized to a non-constant speed of sound in the same way as discussed in Section~\ref{sec:ns_bjorken_qcd}.

\begin{figure}[tb]
	\centering
	\includegraphics[width=0.5\textwidth]{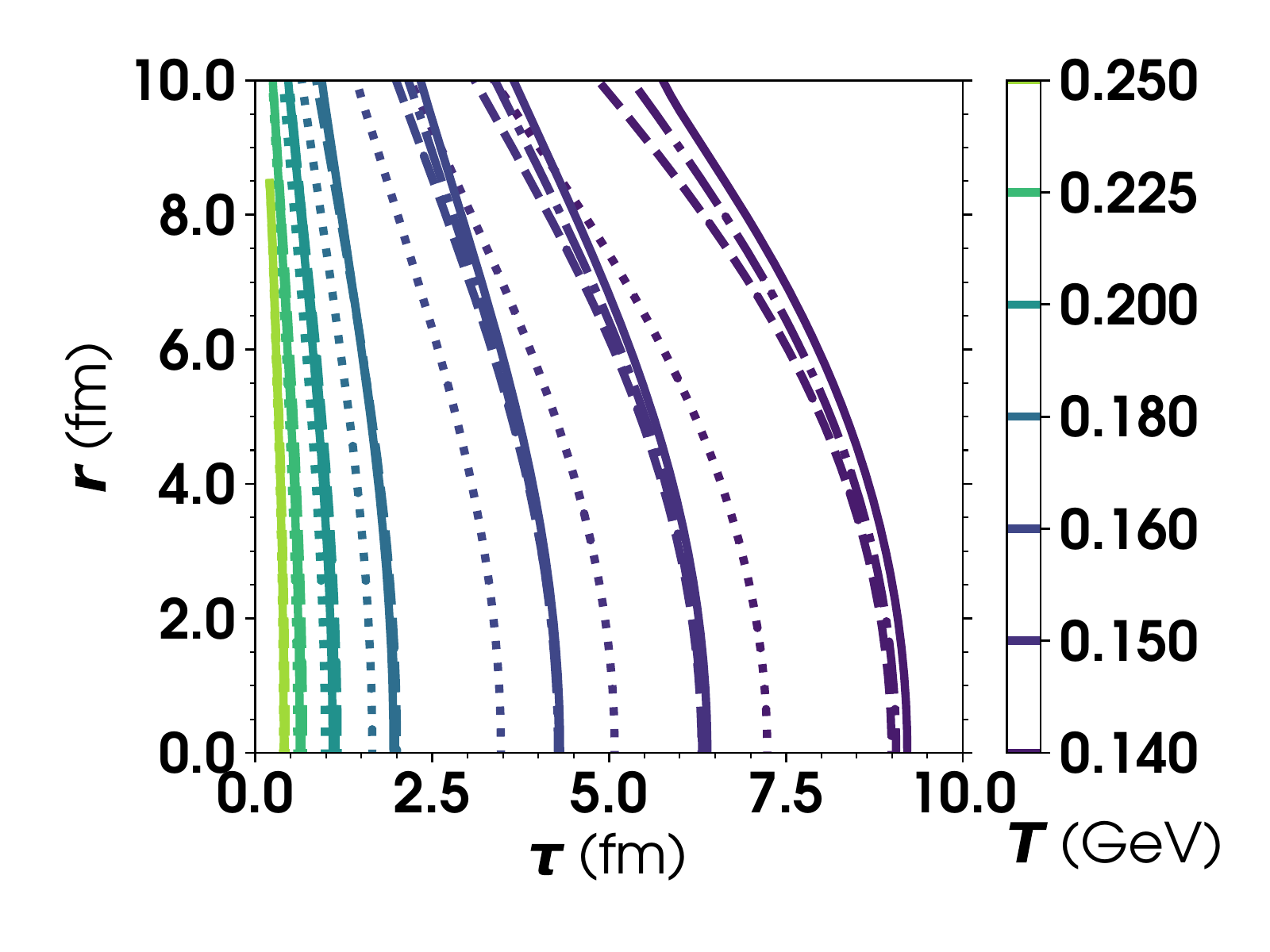}
	(a)
	\includegraphics[width=0.5\textwidth]{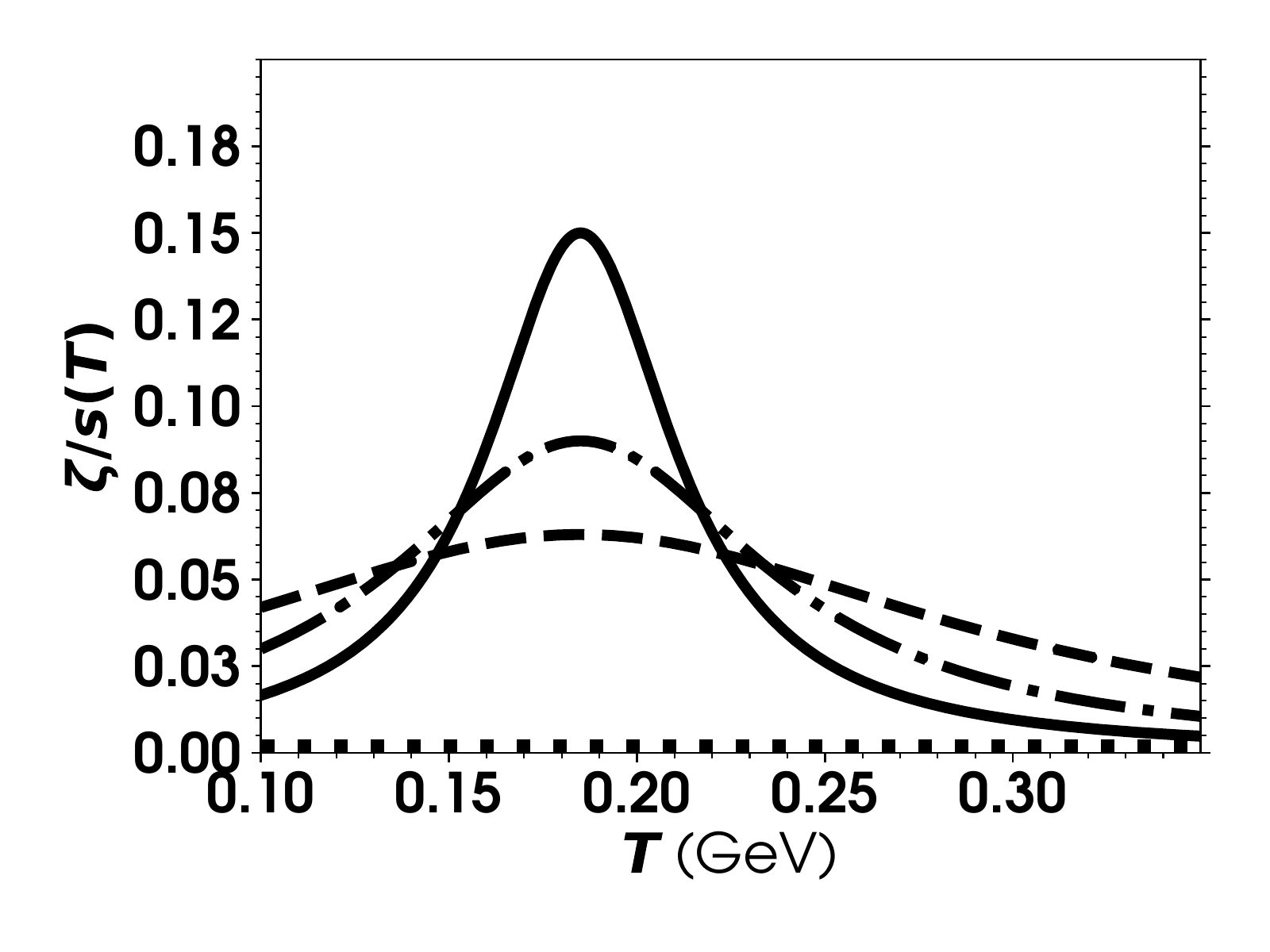}
	(b)
	\caption{
		(a) Temperature profile obtained from a Gaussian initial condition with $T_0=300$~MeV, $\sigma=20$~fm and no initial transverse flow, for the (b) three different temperature-dependent $\zeta/s(T)$ with equivalent Bjorken effective bulk viscosities at $r=0$~fm. The thinner dotted line is the ideal result.
	}
	\label{fig:ns_beyond_bjorken_equiv_bulk_sigma20_naive}
\end{figure}

The approximate effective viscosity of each point in $r$ is given by
\begin{equation}
\langle \zeta/s \rangle_{\textrm{eff}}(r) \approx \frac{\int_{T_f}^{T_0(r)}  d T^\prime \left(\frac{T^\prime}{T_0(r)}\right)^{c_s^{-2}\left(\sqrt{T_0(r) T^\prime}\right)-2} \zeta/s(T^\prime) }{ \int_{T_f}^{T_0(r)}  d T^\prime \left(\frac{T^\prime}{T_0(r)}\right)^{c_s^{-2}\left(\sqrt{T_0(r) T^\prime}\right)-2}  } \; .
\label{eq:Veff_bjorken_qcd_cylindrical}
\end{equation}
We see that, in theory, the effective viscosity will be different at each point: if two $\zeta/s(T)$ are chosen to have the same effective viscosity at a given $r$, they will likely have a different effective viscosity at other values of $r$.
This difference in effective viscosities will depend on the size of the transverse gradients, controlled here by the initial width $\sigma$.
An example\footnote{Note that the numerical solutions of 1+1D \emph{viscous} relativistic Navier-Stokes hydrodynamics shown in this section were obtained with a second-order viscous relativistic hydrodynamics solver~\cite{Schenke:2010nt,Schenke:2010rr,Paquet:2015lta,Paquet:2015lta} with a relaxation time sufficiently small to converge to the Navier-Stokes (first-order hydrodynamics) result.} is shown in Fig.~\ref{fig:ns_beyond_bjorken_equiv_bulk_sigma20_naive}(a) with $\sigma=20$~fm and $T_0=300$~MeV as initial conditions. Three different parametrization of bulk viscosity, shown in Fig.~\ref{fig:ns_beyond_bjorken_equiv_bulk_sigma20_naive}(b), have been chosen so as to have the same Bjorken effective viscosity at $r=0$. These three parametrization of bulk viscosity, which would lead to essentially indistinguishable hydrodynamics evolution in the limit $\sigma \to \infty$, can be distinguished at late times and larger $r$ in a system which undergoes transverse expansion. The difference between the three $\zeta/s(T)$ is nevertheless small compared to the overall effect of bulk viscosity on the temperature profile.

To study systems with larger transverse expansion, it is necessary to go beyond the regime where Eq.~\ref{eq:tube_solution} hold.
In what follows, we explore in more details the case of a system with cylindrical symmetry, and provide a more general definition of effective viscosity valid at arbitrarily late times.

\subsection{Effective viscosities in a cylindrically-symmetric Bjorken system: general case}

\label{sec:effective_visc_1_plus_1}

Using the method of characteristics, the equation of motion for temperature in a cylindrically-symmetric boost-invariant system (Eq.~\ref{eq:NS_T_cyl_eom}) can be rewritten as:
\begin{eqnarray}
\frac{d \tau(\chi)}{d \chi}&=&\sqrt{1+u^r\left(\tau(\chi),r(\chi)\right)^2} \;, \nn \\
\frac{d r(\chi)}{d \chi}&=& u^r\left(\tau(\chi),r(\chi)\right) \;, \nn \\
\frac{d \ln T(\chi)}{d \chi}&=& -c_s^2(T(\chi)) \theta\left(\tau(\chi),r(\chi)\right) \left[ 1 - \frac{\theta}{T} \frac{\zeta}{s}(T) \right] \;, \nn \\
\tau(\chi=0)&=&\tau_0 \;, \nn \\
r(\chi=0)&=&r_0 \;.
\label{eq:cyl_hydro_charac}
\end{eqnarray}

\begin{figure}[tb]
	\centering
	\includegraphics[width=0.5\textwidth]{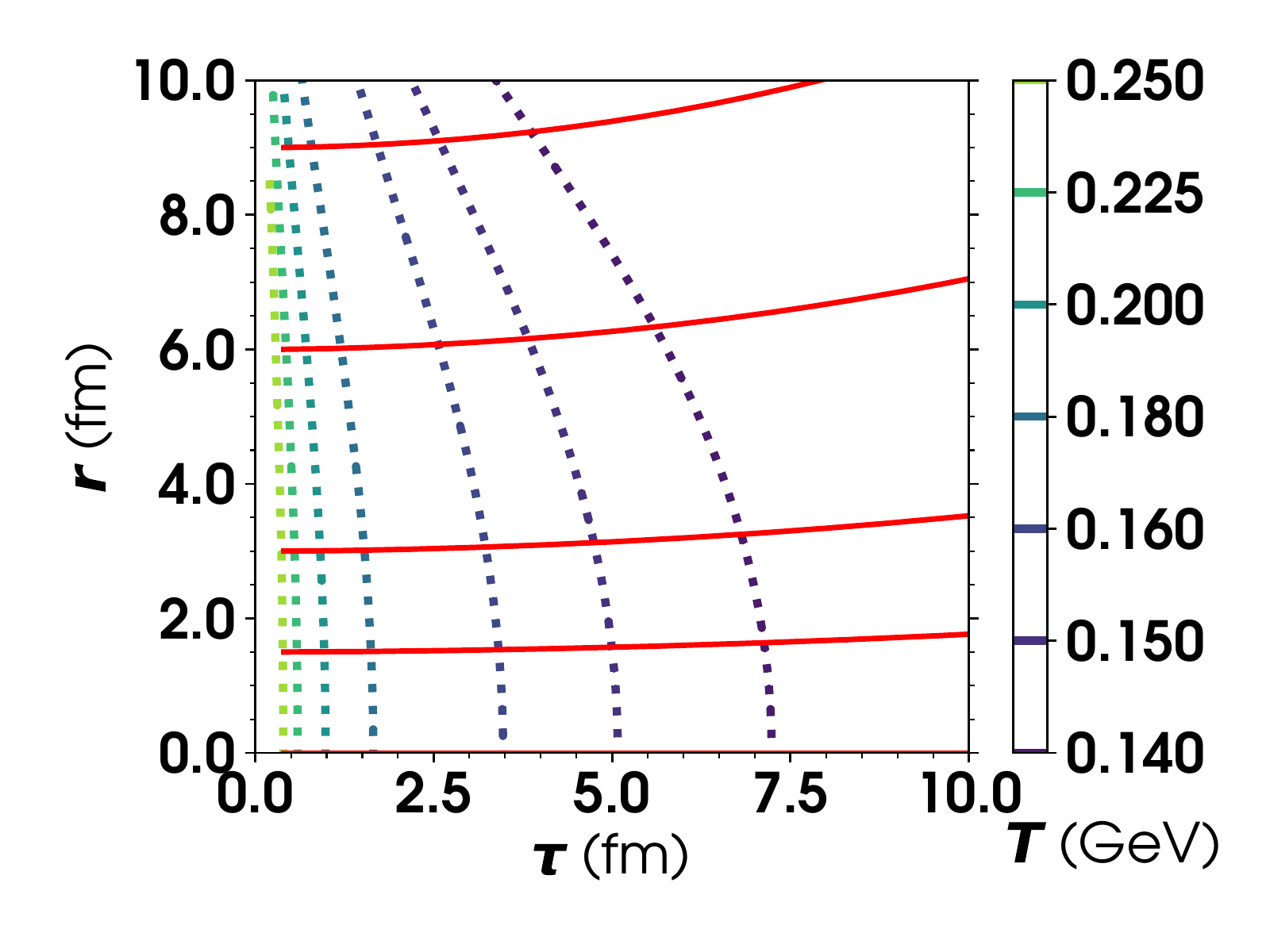}
	(a)
	\includegraphics[width=0.5\textwidth]{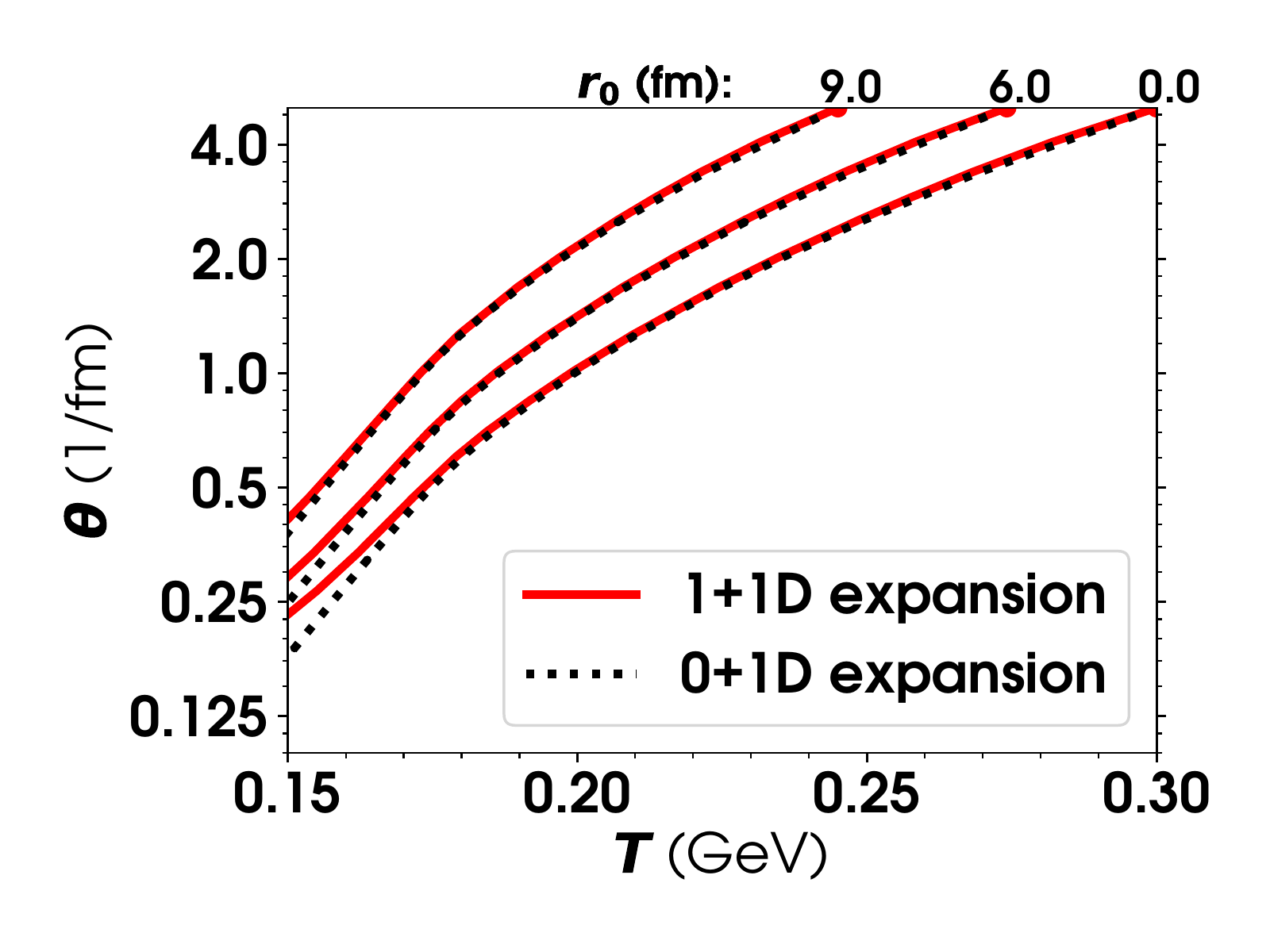}
	(b)
	\caption{
		(a) Ideal temperature profile and characteristic curves obtained from a Gaussian initial condition with $T_0=300$~MeV, $\sigma=20$~fm, $\tau_0=0.2$~fm and no initial transverse flow, and (b) trajectories of a subset of characteristic curves in the $\theta-T$ space. For reference, to illustrate the effect of the transverse expansion, the $\theta-T$ trajectories for $0+1$D expansion are shown with dashed lines in (b).
	}
	\label{fig:ideal_characteristics_sigma20}
\end{figure}

For a given $r_0$, the equations of motion for $\tau(\chi)$ and $r(\chi)$ form characteristic curves. For reference, these characteristic curves are illustrated in Fig.~\ref{fig:ideal_characteristics_sigma20}(a) for the ideal case with $T_0=300$~MeV, $\sigma=20$~fm and $\tau_0=0.2$~fm (same parameters as used earlier in this section). Each characteristic follow a different trajectory in expansion rate and temperature (``$\theta-T$ space''), illustrated on Fig.~\ref{fig:ideal_characteristics_sigma20}(b). This did not happen in 0+1D, where all points in the transverse direction had the same trajectory in $\theta-T$. In what follows, we discuss a more general definition of effective viscosity and its close relationship with the $\theta-T$ trajectories illustrated in Fig.~\ref{fig:ideal_characteristics_sigma20}(b).

\subsubsection{Central transverse position ($r_0=0$)}

By symmetry, at the center of the fluid ($r_0=0$), $u^r=0$, and the characteristic is $\chi=\tau$:
\begin{equation}
\frac{d \ln T}{d \tau}= -c_s^2(T) \theta\left(\tau,r=0\right) \left[ 1 - \frac{\theta}{T} \frac{\zeta}{s}(T) \right] \; .
\end{equation}
The effective viscosity can be defined the same way as in Section~\ref{sec:ns_bjorken_conf}:
\begin{equation}
\langle \zeta/s \rangle_{\textrm{eff}}(r=0)\approx \frac{\int d\tau c_s^2(T) \theta\left(\tau,r=0\right) \frac{\theta}{T} \frac{\zeta}{s}(T)}{\int d\tau c_s^2(T) \theta\left(\tau,r=0\right) \frac{\theta}{T}} \; .
\label{eq:Veff_cylindrical_center}
\end{equation}

The difference with Section~\ref{sec:ns_bjorken_conf} is that the expansion rate is not $\theta=1/\tau$ anymore\footnote{For reference, at $r=0$, the ideal expansion rate $\theta$ can be approximated by
	\be
	\theta(\tau,r=0)\approx \frac{1}{\tau}+\frac{4 \left(\tau - \tau_0 \left(\frac{\tau}{\tau_0}\right)^{c_s^2}\right)}{(1-c_s^2) \sigma^2 \left(1 + \frac{2 (\tau-\tau_0)^2}{(1+c_s^2) \sigma^2}\right)}
	\label{eq:theta_cyl_r0}
	\ee
where $c_s^2$ is assumed to be a constant. Because Eq.~\ref{eq:theta_cyl_r0} depends weakly on the speed of sound, it remains a good approximation when the QCD equation of state is used, as long as $c_s^2$ is taken in a reasonable range of values ($c_s^{-2}\sim 3-5$; see Appendix~\ref{sec:appendix_bjorken} for a discussion of preferable constant values of $c_s^{-2}$ to use in such instances).}, a consequence of the $\partial_r u^r$ term in Eq.~\ref{eq:theta_cyl}.
Equation~\ref{eq:Veff_bjorken_qcd_cylindrical} will only provide a good approximation of Eq.~\ref{eq:Veff_cylindrical_center} if the effect of viscosity is concentrated at very early times (that is, only if the support of $\zeta/s(T)$ is close to the initial temperature of the fluid at $r_0=0$).

\subsubsection{General transverse position $r$}

In general, an effective viscosity can be defined for each characteristic, and can be labelled by the initial value of $r_0=r(\chi=0)$:
\begin{equation}
\langle \zeta/s \rangle_{\textrm{eff}}(r_0)\approx \frac{\int d\chi c_s^2(T) \theta\left(\tau(\chi),r(\chi)\right) \frac{\theta}{T} \frac{\zeta}{s}(T)}{\int d\chi c_s^2(T) \theta\left(\tau(\chi),r(\chi)\right) \frac{\theta}{T}} \; .
\label{eq:Veff_cylindrical_gen}
\end{equation}

\begin{figure}[tb]
	\centering
	\includegraphics[width=0.5\textwidth]{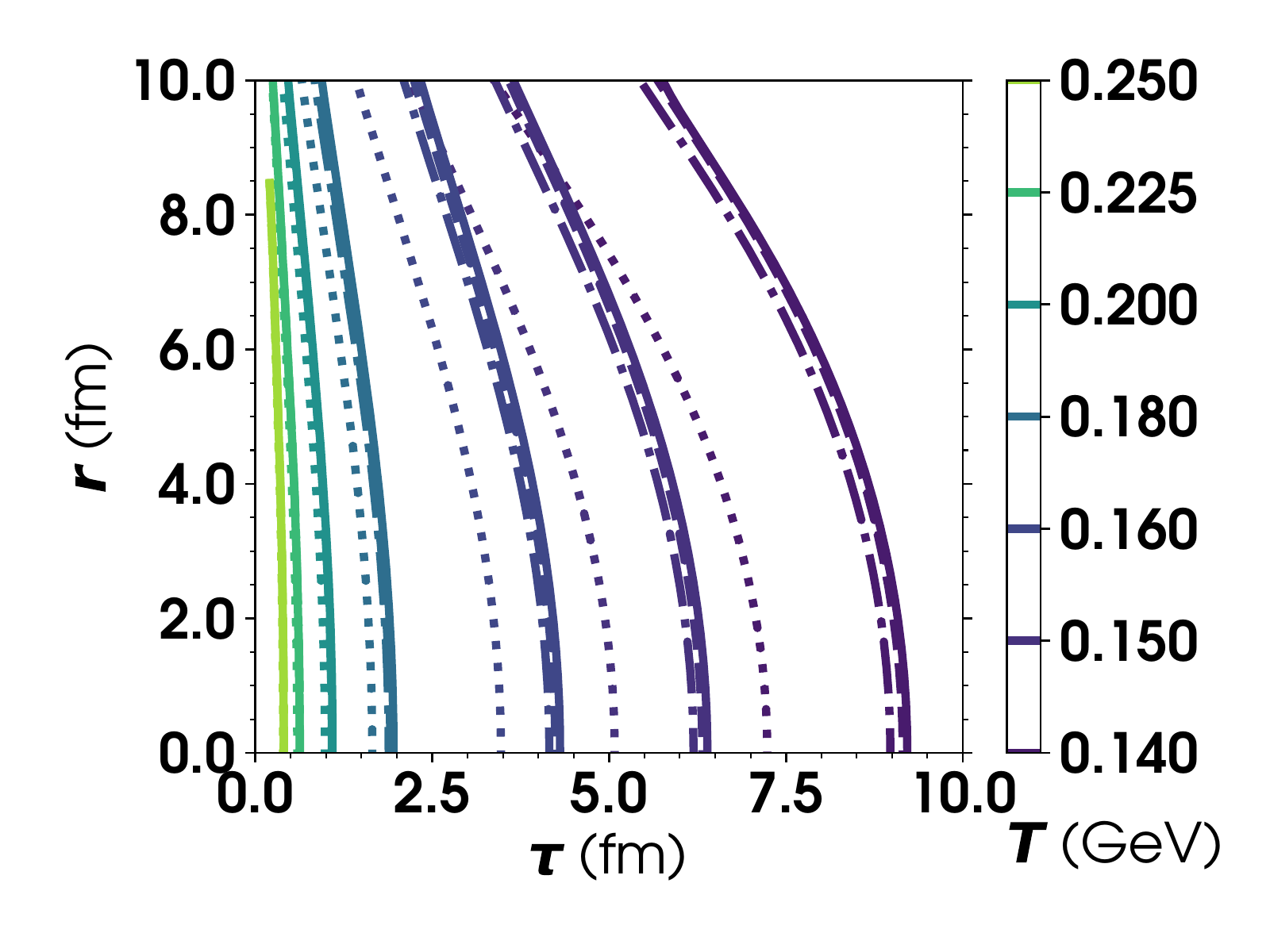}
	(a)
	\includegraphics[width=0.5\textwidth]{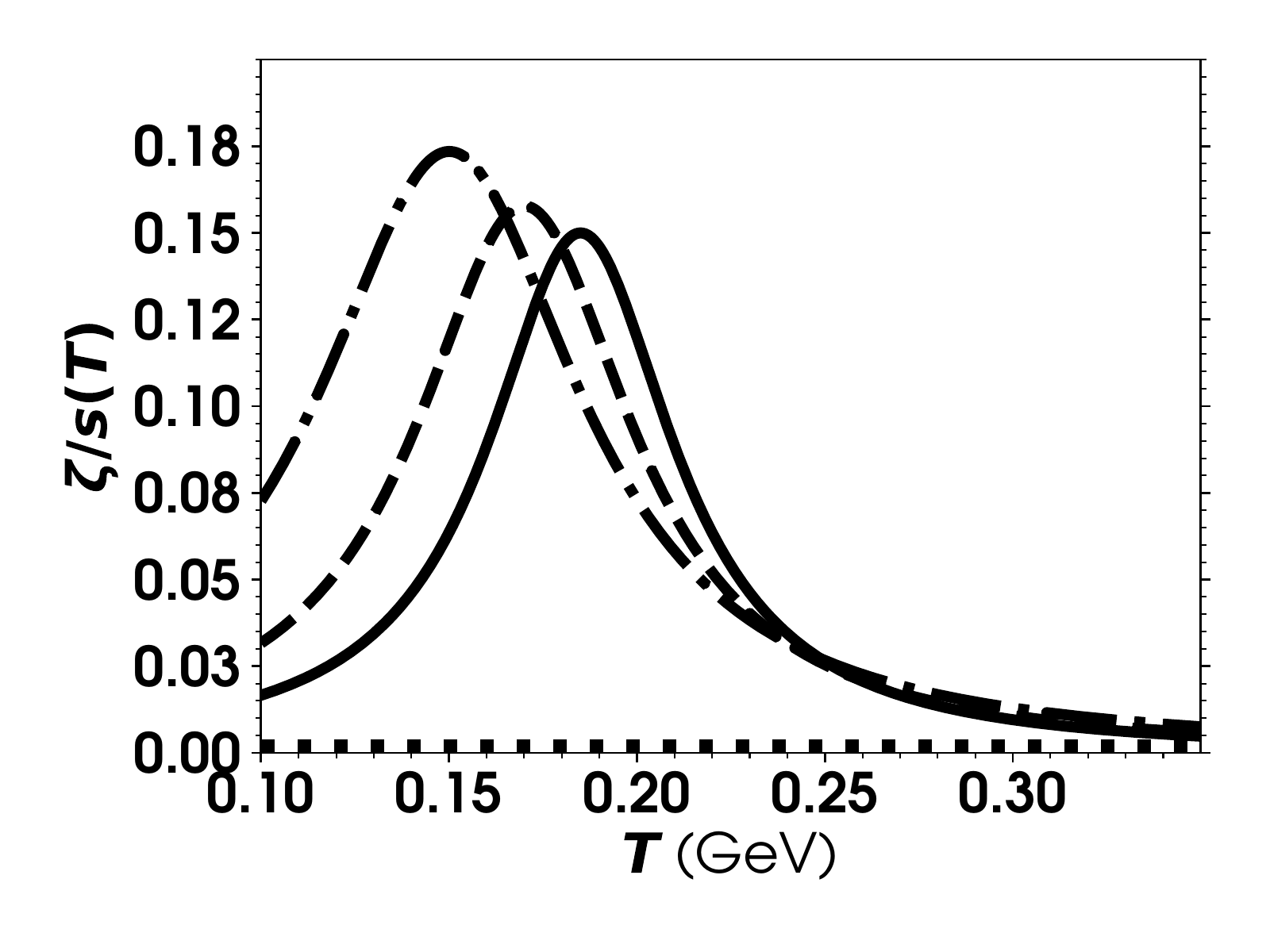}
	(b)
	\caption{
		(a) Temperature profile obtained from a Gaussian initial condition with $T_0=300$~MeV, $\sigma=20$~fm and no initial transverse flow, for the (b) three different temperature-dependent $\zeta/s(T)$ with equivalent Bjorken effective bulk viscosities at $r_0=0$ and $10$~fm. The thinner dotted line is the ideal result.
	}
	\label{fig:ns_beyond_bjorken_equiv_bulk_sigma20_three_params}
\end{figure}

In the Bjorken case, it was sufficient to insure that different parametrizations of $\zeta/s(T)$ have the same effective viscosity at a single value of $r_0$. In the present case, if $\zeta_1/s(T)$ is one parametrization and $\zeta_2/s(T)$ is a second one, one must ensure
\begin{equation}
\langle \zeta_1/s(T) \rangle_{\textrm{eff}}(r_0) = \langle \zeta_2/s(T) \rangle_{\textrm{eff}}(r_0) \; \forall r_0 \; .
\label{eq:r0_dep_eff_visc}
\end{equation}
In practice, the temperature profile is smooth, 
and a discrete set of $r_0$ should be sufficient.
For example, recall the example shown in Figure~\ref{fig:ns_beyond_bjorken_equiv_bulk_sigma20_naive}.
Using Eq.~\ref{eq:Veff_cylindrical_gen} for only two values of $r_0$, $0$ and $10$~fm, results in Fig.~\ref{fig:ns_beyond_bjorken_equiv_bulk_sigma20_three_params}: three different parametrizations of $\zeta/s(T)$ that leads to a very similar temperature profiles for a wide range of transverse positions $r$.

These parametrizations are obtained using the following steps:
\begin{itemize}
	\item Solve numerically the \emph{ideal} hydrodynamic equations to obtain the ideal profiles for the temperature $T_I(\tau,r)$, the flow velocity $u_I^r(\tau)$ and the expansion rate $\theta_I(\tau,r)$.
	\item Use $u_I^r(\tau)$ to find numerically the ideal characteristic solutions for $\tau(\chi)$ and $r(\chi)$ 
	\item Minimize Eq.~\ref{eq:r0_dep_eff_visc} with respect to two different parametrizations of $\zeta/s(T)$, for a discrete set of $r_0$.
\end{itemize}

These numerical steps\footnote{The ideal cylindrical relativistic Navier-Stokes equations can be solved with Mathematica's ``NDSolve'' function. The characteristics can be calculated the same way.} are necessary since the function $\theta(\tau,r)$ is not known for a cylindrically-symmetric boost-invariant fluid. In the earlier 0+1D invariant case, $\theta(\tau)$ was simply $1/\tau$.

Because this process involves minimizing a function [$\zeta/s(T)$] over a range of temperature, there can be a wide variety of solutions, depending on the constraints imposed on the functional form of  $\zeta/s(T)$.
The minimization procedure is highly non-linear, and significant changes in $\zeta/s(T)$ can be necessary to obtain slightly better agreement in the temperature profiles.
This can be seen clearly in Fig~\ref{fig:ns_beyond_bjorken_equiv_bulk_sigma20_naive} and Fig.~\ref{fig:ns_beyond_bjorken_equiv_bulk_sigma20_three_params}, where the better agreement of the temperature profiles at large $r$ in Fig.~\ref{fig:ns_beyond_bjorken_equiv_bulk_sigma20_three_params} was obtained by completely changing parametrizations of $\zeta/s(T)$ used in Fig~\ref{fig:ns_beyond_bjorken_equiv_bulk_sigma20_naive}.

In theory, one could find equivalent parametrizations of $\zeta/s(T)$ by brute force numerical analysis: solve the equations of \emph{viscous} hydrodynamics numerically with a large ensemble of $\zeta/s(T)$ and evaluate numerically which ones have similar hydrodynamic evolutions, as quantified (for example) by their temperature profile.\footnote{Pushing this one step forward, and trying to identify numerically which parametrizations of $\zeta/s(T)$ lead to the same final distribution of hadrons in a realistic heavy ion collision would essentially be the same as current Bayesian analysis such as Ref.~\cite{Novak:2013bqa,Bernhard:2016tnd,Bernhard:2019bmu}.} This is not the approach we are putting forward. Equation~\ref{eq:Veff_cylindrical_gen} only needs the ideal solution to the hydrodynamic equations, which is in general straightforward to obtain numerically. More importantly, Eq.~\ref{eq:Veff_cylindrical_gen} provides intuition on the relation between $\zeta/s(T)$, the temperature profile, the expansion rate and the resulting effect of viscosity. We illustrate this important point in the next section.

\begin{figure}[tb]
	\centering
	\includegraphics[width=0.5\textwidth]{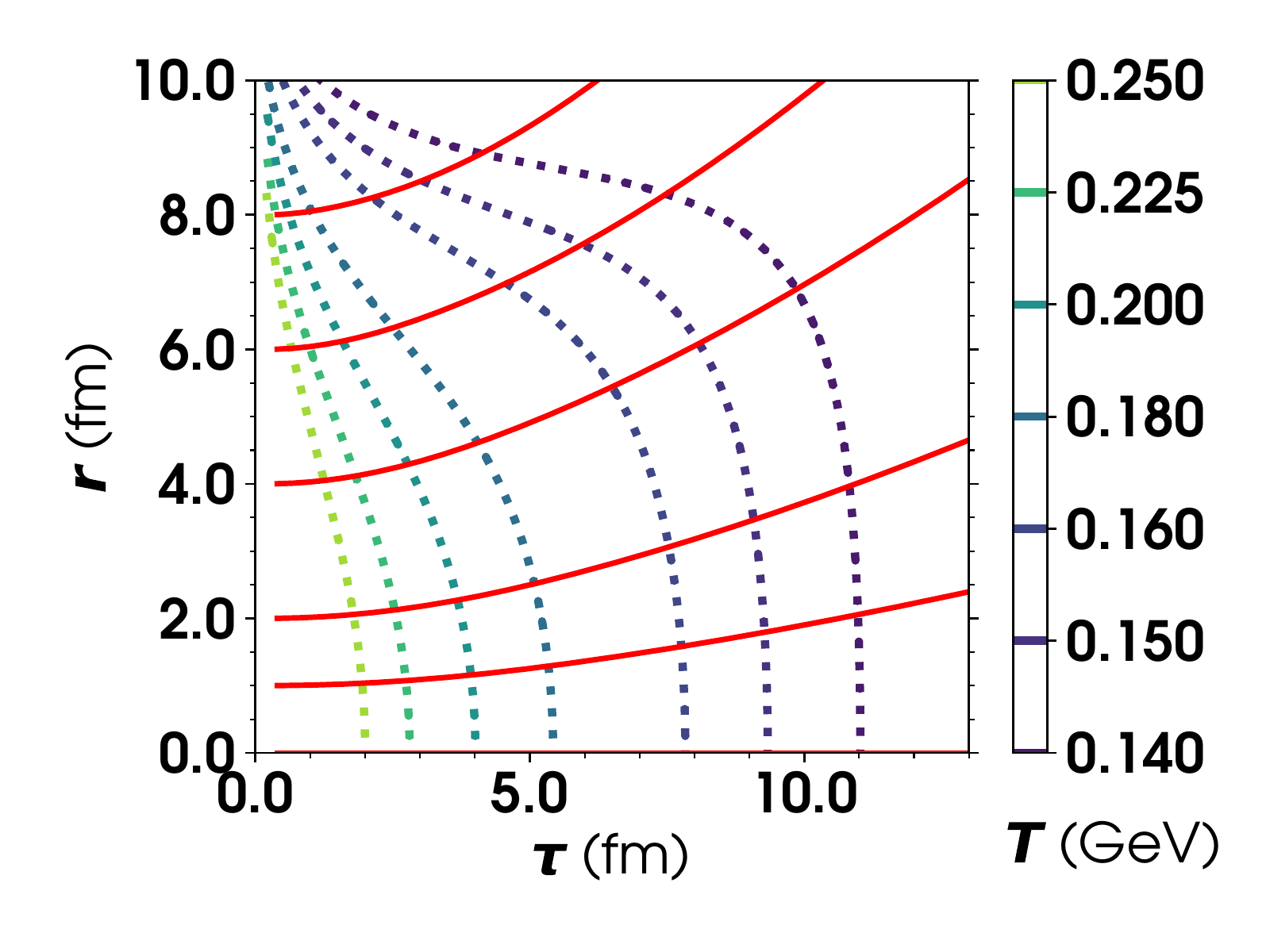}
	(a)
	\includegraphics[width=0.5\textwidth]{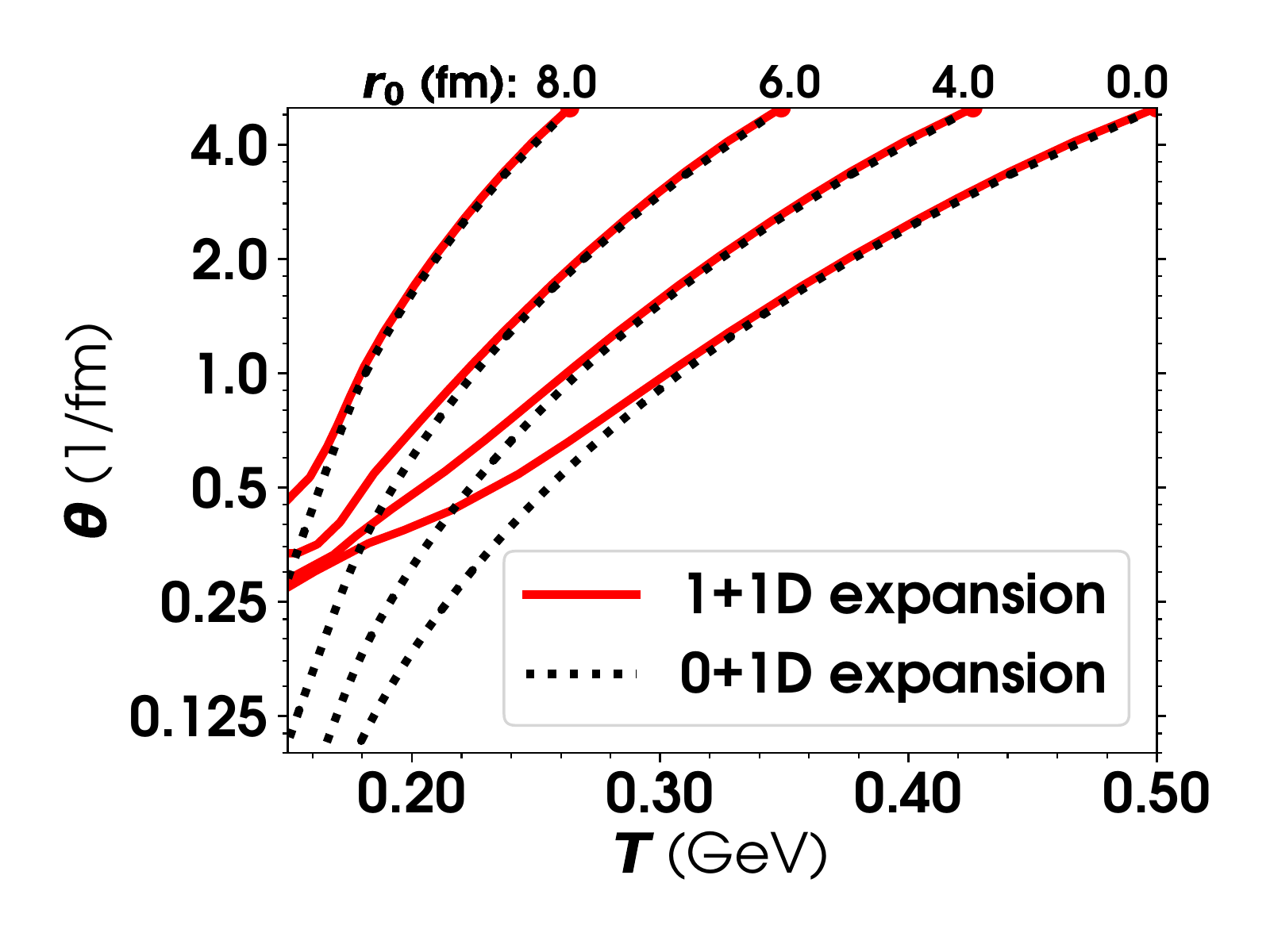}
	(b)
	\caption{
		(a) Ideal temperature profile and characteristic curves obtained from a Gaussian initial condition with $T_0=300$~MeV, $\sigma=10$~fm, $\tau_0=0.2$~fm and no initial transverse flow, and (b) trajectories of a subset of characteristic curves in the $\theta-T$ space. The $\theta-T$ trajectories for $0+1$D expansion are shown with dashed lines in (b).
	}
	\label{fig:ideal_characteristics_sigma10}
\end{figure}

\subsubsection{Larger transverse gradients}

Suppose the width of the initial Gaussian temperature profile is reduced from the $\sigma=20$~fm used above to $\sigma=10$~fm; this increases the size of the transverse gradients, making the effect of the transverse expansion more visible. We increase the initial temperature at the center to $T_0=500$~MeV, so that the evolution covers a similar range of temperature as the previous example. The ideal temperature profile and characteristic curves are shown on Fig.~\ref{fig:ideal_characteristics_sigma10}.  The expansion rate varies significantly along characteristics in the $\theta-T$ plane, meaning that the effect of viscosity on the hydrodynamic evolution will vary significantly across the transverse plane. From  Fig.~\ref{fig:ideal_characteristics_sigma10}, we can see that the expansion rate along characteristics, $\theta(\tau(\chi),r(\chi))$, can be written as $\theta(T,r_0)$.
That is, given an initial transverse position, the expansion rate along a characteristic can be expressed as a function of temperature alone. 
This allows Equation~\ref{eq:Veff_cylindrical_gen} to be rewritten:
\begin{equation}
\langle \zeta/s \rangle_{\textrm{eff}}(r_0)\approx \frac{\int \frac{d T^\prime}{T^\prime}  \frac{\theta(T^\prime,r_0)}{T^\prime} \frac{\zeta}{s}(T^\prime)}{\int \frac{d T^\prime}{T^\prime}  \frac{\theta(T^\prime,r_0)}{T^\prime}}
\label{eq:Veff_cylindrical_T}
\end{equation}
with $\theta(T^\prime,r_0)$ being the function shown in Fig.~\ref{fig:ideal_characteristics_sigma10}(b); at the moment, this function is only known numerically. The range of integration is from the final temperature $T_f(r)$ to the initial temperature $T_0(r)$.

\begin{figure}
	\centering
	\includegraphics[width=0.5\textwidth]{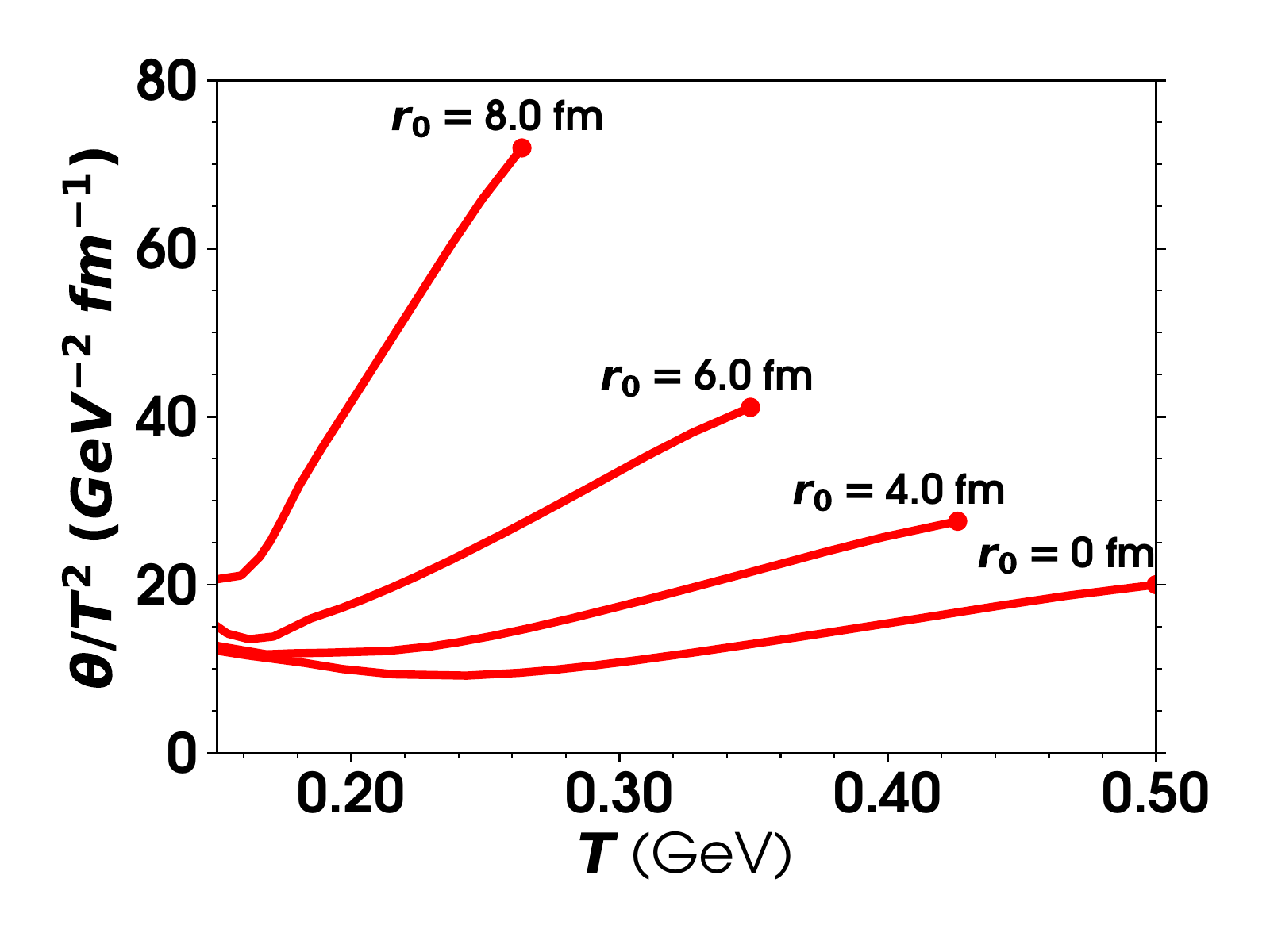}
	\caption{Product $\theta(T,r_0)/T^2$ that weights $\zeta/s(T)$ when evaluating the effective viscosity at a given $r_0$ in Eq.~\ref{eq:Veff_cylindrical_T}. Four different characteristics are shown, $r_0=0,4,6 \; \& \; 8$~fm.}
	\label{fig:theta_over_T2_ideal_characteristics_sigma10}
\end{figure}

The form of Eq.~\ref{eq:Veff_cylindrical_T} allows an easier visualization of the constraints on $\zeta/s(T)$ from the different positions in the transverse plane.
For a given point in the transverse plane (that is, for a given value of $r_0$), the weight that multiplies the different parametrizations of $\zeta/s(T)$ is $\theta(T,r_0)/T^2$, illustrated in Fig.~\ref{fig:theta_over_T2_ideal_characteristics_sigma10}. Different $\zeta/s(T)$ will lead to similar temperature profiles if the area under the product $[\zeta/s(T)] \theta(T,r_0)/T^2$ remains the same for all $r_0$.

\begin{figure}[tb]
	\centering
	\includegraphics[width=0.5\textwidth]{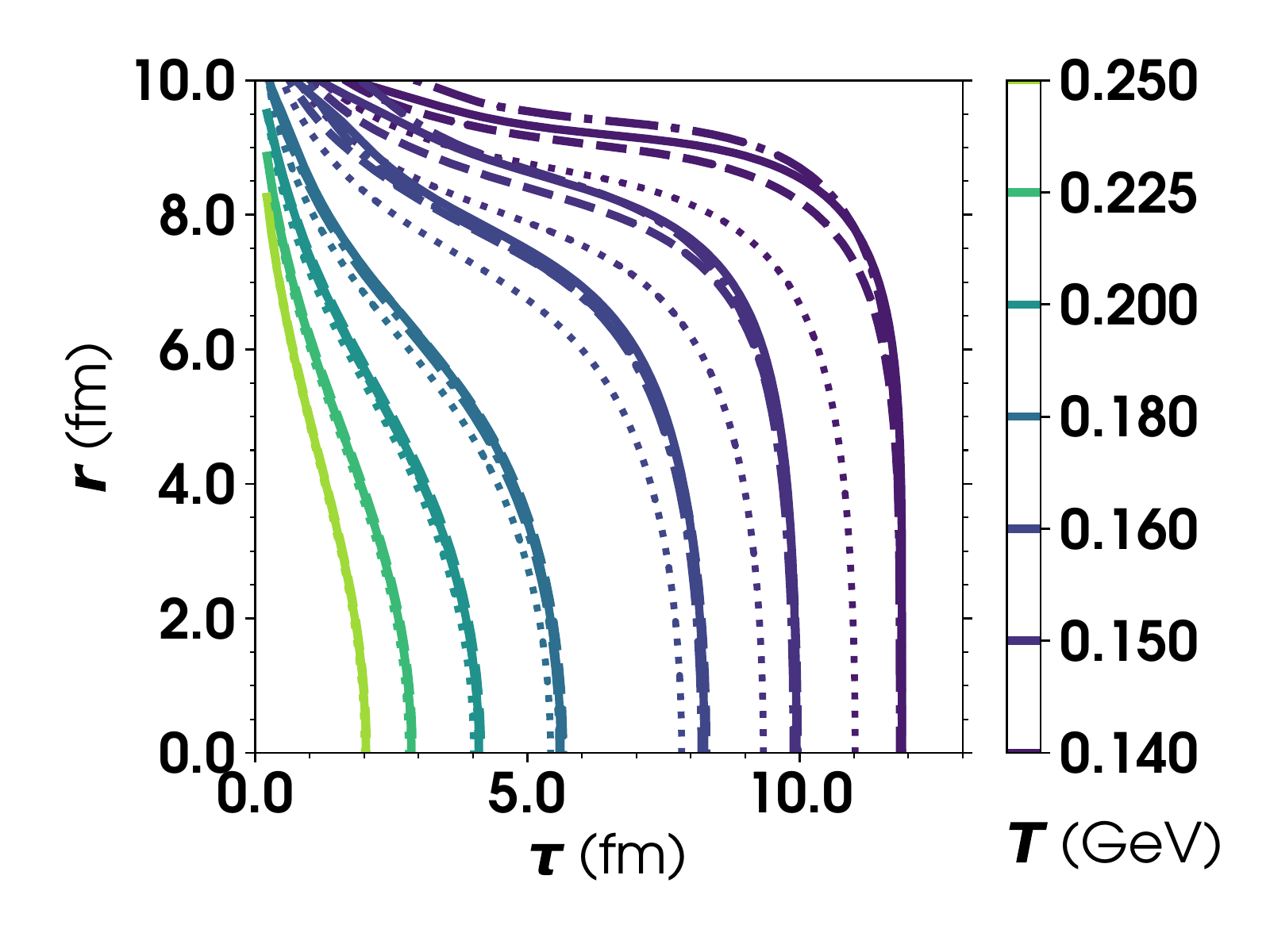}
	(a)
	\includegraphics[width=0.5\textwidth]{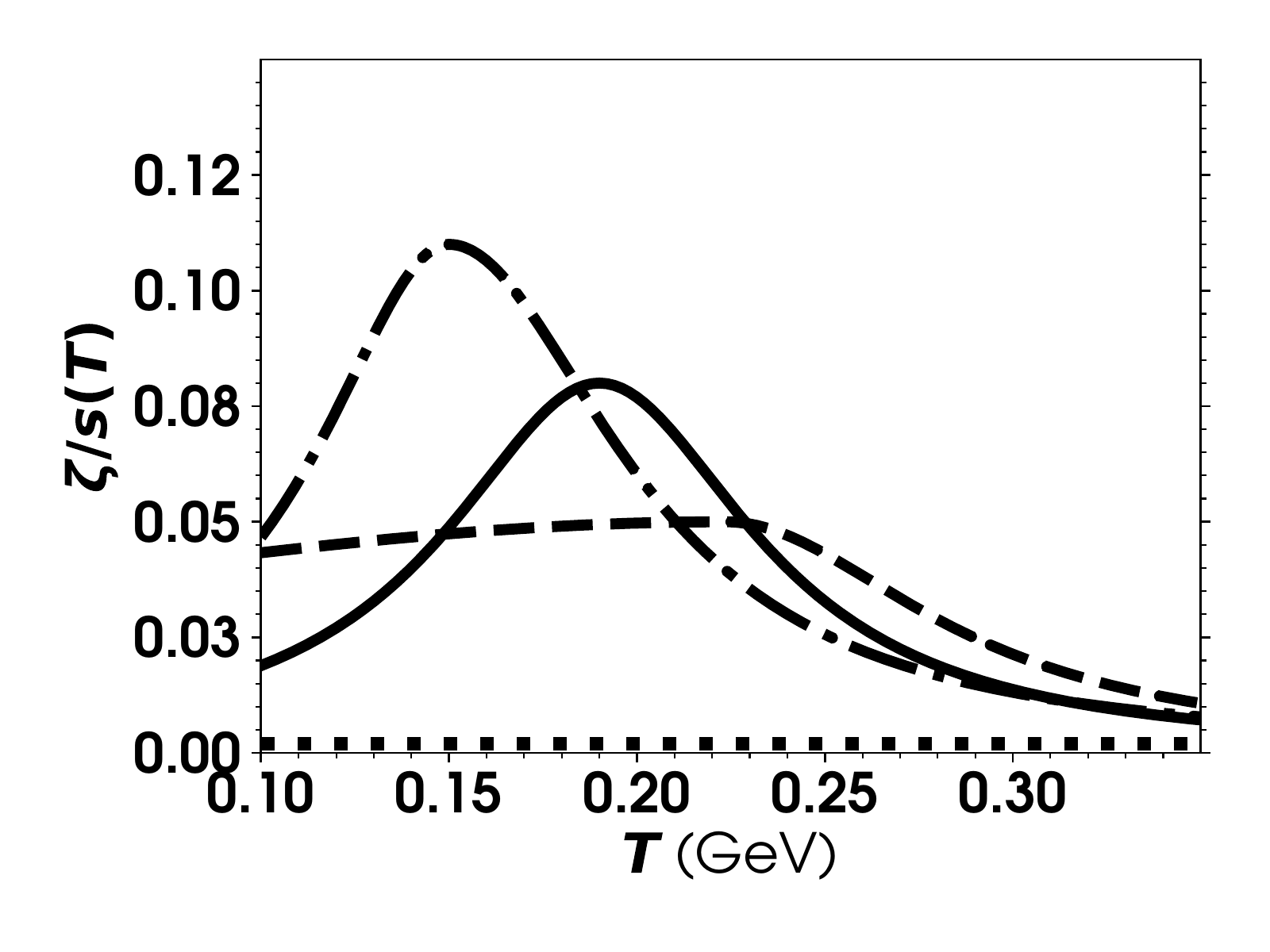}
	(b)
	\caption{
		(a) Temperature profile obtained from a Gaussian initial condition with $T_0=500$~MeV, $\sigma=10$~fm and no initial transverse flow, for the (b) three different temperature-dependent $\zeta/s(T)$ with similar Bjorken effective bulk viscosities across a wide range of $r_0$ between $0$ and $10$~fm. The thinner dotted line is the ideal result.
	}
	\label{fig:ns_beyond_bjorken_equiv_bulk_sigma10}
\end{figure}

Choosing $T_f(r)$ to be a constant, $T_f=150$~MeV, three such different parametrizations of $\zeta/s(T)$ that lead to similar hydrodynamic evolution are shown in Fig.~\ref{fig:ns_beyond_bjorken_equiv_bulk_sigma10}. 
They were obtained by minimizing Eq.~\ref{eq:Veff_cylindrical_T}, with $\theta(T^\prime,r_0)$ having been obtained numerically.
This time, a more flexible parametrization of $\zeta/s(T)$ was used, with the peak allowed to be asymmetric. This still represents a very small subset of the space of functions that could be minimized with Eq.~\ref{eq:Veff_cylindrical_T}. Again, the temperature profiles are not identical, but are very similar given the significant differences in the temperature dependence of $\zeta/s(T)$, as well as  given how much they deviate from the ideal solution (shown with dashed lines).

\begin{figure}
	\centering
	\includegraphics[width=0.5\textwidth]{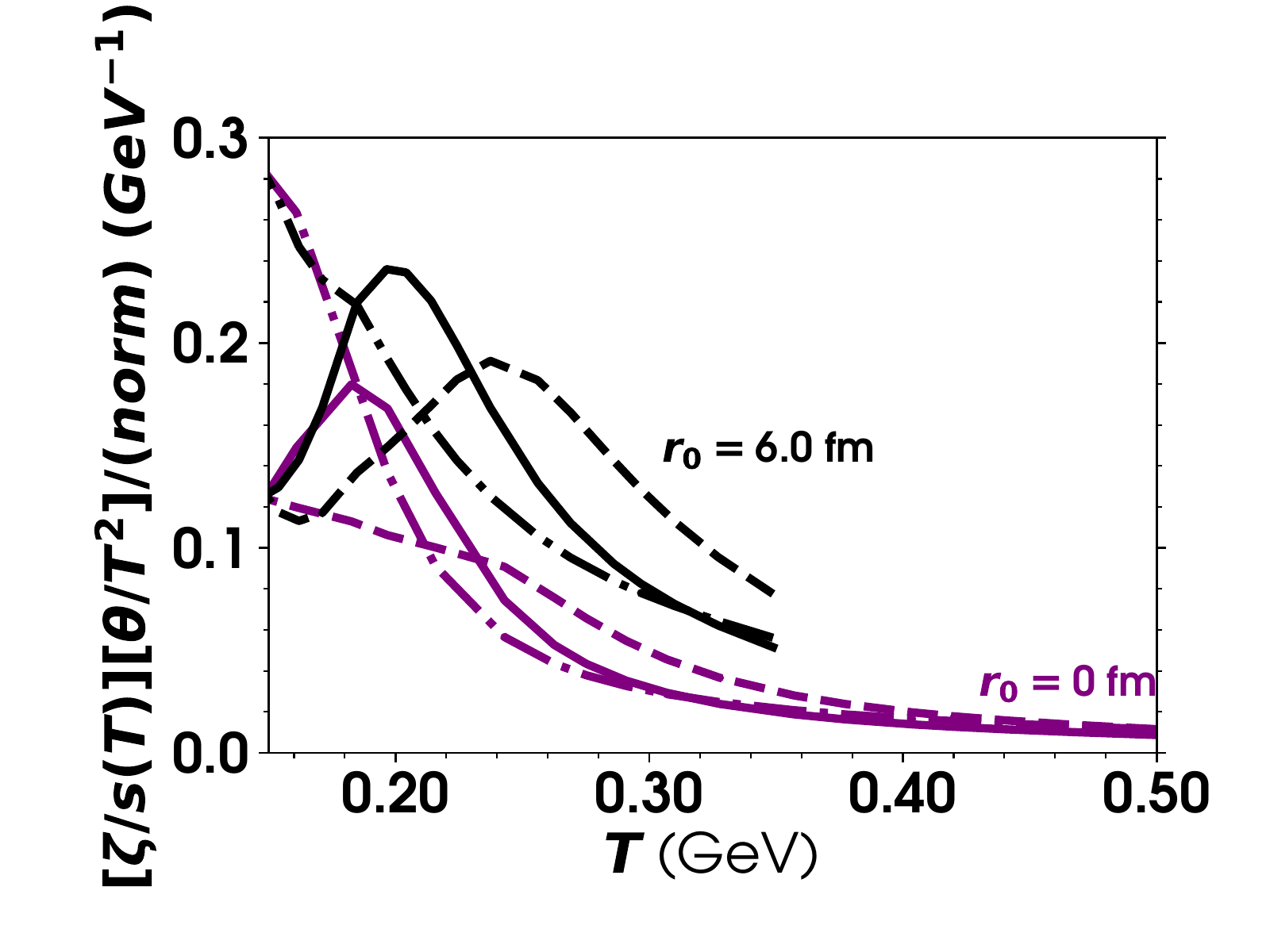}
	\caption{Product $\zeta/s(T) \; \theta(T,r_0)/T^2$ from Eq.~\ref{eq:Veff_cylindrical_T}, normalized such that the area under the curves represent their effective viscosity. Shown for two characteristics with $r_0=0$ and $r_0=6$~fm.}
	\label{fig:zetaovers_theta_over_T2_ideal_characteristics_sigma10}
\end{figure}

Figure~\ref{fig:zetaovers_theta_over_T2_ideal_characteristics_sigma10} shows the product $[\zeta/s(T)] \theta(T,r_0)/T^2$ for the three parametrizations shown in Fig.~\ref{fig:ns_beyond_bjorken_equiv_bulk_sigma10}(b). It is normalized by the denominator of Eq.~\ref{eq:Veff_cylindrical_T}, such that the area under each curve is its effective viscosity.
Only two characteristics, $r_0=0$ and $6$~fm, are shown for clarity. The effective viscosity is approximately $0.02$ for $r_0=0$~fm, while it is larger than to $0.03$ for $r_0=6$~fm.
As highlighted by Fig.~\ref{fig:zetaovers_theta_over_T2_ideal_characteristics_sigma10}, these values are the result of the non-trivial dependence of the expansion rate with $\tau$ and $r$ combined with the exact form of $\zeta/s(T)$.
Importantly, as should be clear by now, no single \emph{constant} value of $\zeta/s$ should be expected to provide a good approximation of the parametrizations of $\zeta/s(T)$ shown in Fig.~\ref{fig:ns_beyond_bjorken_equiv_bulk_sigma10}(b): one clearly needs a larger value of effective viscosity at large $r$ than at small $r$. At best, a compromise could be found between the effective viscosities favored by large and small $r$, an effective viscosity that would be somewhere between $0.02$ and $0.03$ for the example shown in Fig.~\ref{fig:ns_beyond_bjorken_equiv_bulk_sigma10}.

It must be emphasized that the discussion from this section can be generalized to shear viscosity as well. One of the reasons bulk viscosity was discussed instead of shear viscosity is the slightly simpler form of its equation of motion. The other reason is related to the physics of heavy ion collisions and of the bulk viscosity of QCD. The bulk viscosity of QCD is expected to have a more limited support in temperature than the shear viscosity: while $\eta/s$ of QCD is expected to take values $\gtrsim 0.1$ at most temperatures above $150$~MeV, it is possible that $\zeta/s$ only take non-negligible values for temperature below $\sim 250$~MeV. At these lower temperature, hydrodynamic simulations of heavy ion collisions indicate that the temperature profile is much more uniform than at high temperature. In head-on central collisions of nuclei, the cylindrically-symmetric boost-invariant fluid discussed in this section could be a reasonable approximation of the late-time temperature profile. This could make the results derived in this section more relevant for bulk viscosity than for shear viscosity.

\begin{figure}
	\centering
	\includegraphics[width=0.5\textwidth]{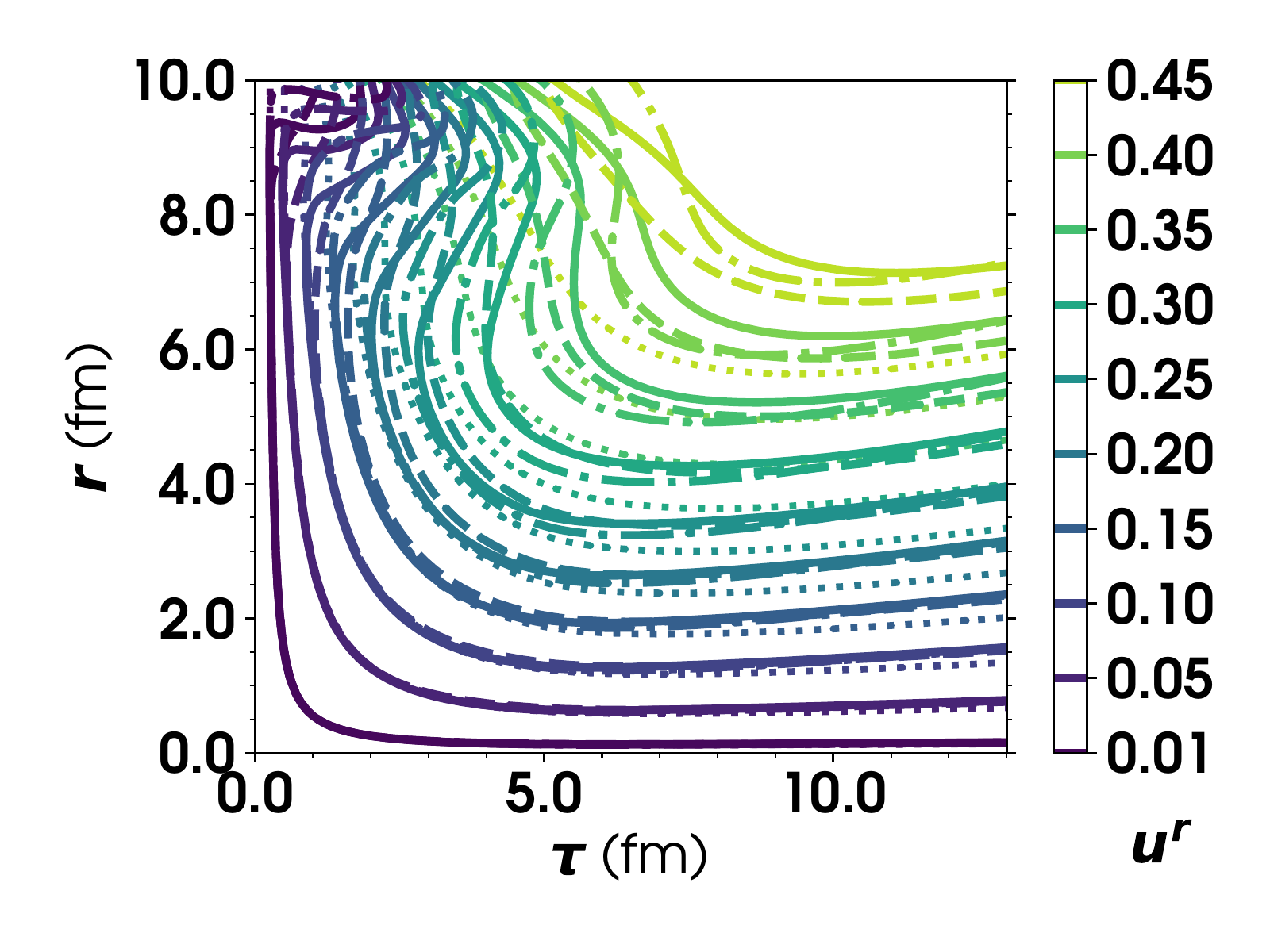}
	\caption{Radial flow velocity corresponding to the temperature profiles shown in Fig.~\ref{fig:ns_beyond_bjorken_equiv_bulk_sigma10}.}
	\label{fig:ur_tau_r_countour_sigma10}
\end{figure}

We conclude this discussion by noting that we have focused solely on the temperature profile of the fluid to study the effective viscosity of the fluid. As we have seen, it does provide a definition of effective viscosity that works reasonably well in practice. Of course, in 1+1D, the flow velocity is also important. This radial flow profile is shown in Fig.~\ref{fig:ur_tau_r_countour_sigma10}, for the example with $\sigma=10$~fm discussed in this section. The flow velocity profile is actually relatively similar for the different parametrization of $\zeta/s(T)$, although differences appear to be larger than for the temperature profile.
Including the flow velocity in the definition of effective viscosity may be difficult, given the more complex form of its equation of motion.
Actually, it is unclear if this inclusion is necessary: it appears unlikely that two smooth temperature profiles could be similar over a range of $\tau$ and $r$ with their corresponding flow velocity being very different --- at least in systems as symmetric as these discussed in this work.
Whether this remains true in a full 3+1D system is less clear; this question will need to be revisited in more details for such systems.
\subsection{Defining a global effective viscosity beyond 0+1D}

Equation~\ref{eq:Veff_cylindrical_gen} defines an effective viscosity along the spacetime trajectory of a characteristic. These characteristics can be labeled by their initial position, for example $\tau_0$ and $r_0$ in the cylindrical 1+1D case discussed in the previous section. 
Different temperature-dependent viscosities --- $\zeta_1/s(T)$ and $\zeta_2/s(T)$ --- which have the same effective viscosity (Equation~\ref{eq:Veff_cylindrical_gen}) --- $\langle \zeta/s(T) \rangle_{\textrm{eff}}(r_0)$ --- will have approximately the same temperature at the end point of the $r_0$-labelled-characteristic.

As we saw in the 0+1D case, along a \emph{single} characteristic, there is an infinite number of different $\zeta/s(T)$ and $\eta/s(T)$ that have the same effective viscosity.
On the other hand, different characteristics generally have different effective viscosities, as seen in the example presented in the previous section.
Beyond 0+1D, for two parametrizations of $\zeta/s(T)$ (or $\eta/s(T)$) to lead to similar overall temperature profiles, they must have similar effective viscosities for a wide range of different characteristics, spanning across the fluid. 
We discussed one way of achieving this in the previous section: for a set of different values of the coordinate $r_0$, minimizing the difference between the effective viscosity $\zeta/s \rangle_{\textrm{eff}}(r_0)$ of two different parametrizations of bulk viscosity, $\zeta_1/s(T)$ and $\zeta_2/s(T)$.

In theory, characteristics can be defined in higher dimensions as well, and similar definitions of effective viscosities along characteristics can be obtained. However, beyond 0+1D, it does not seem possible to define a proper \emph{global} effective viscosity for the fluid as a whole. 
Averages over multiple characteristics of their respective effective viscosities are possible, although referring to such averages as ``global'' effective viscosities would obscure the important conclusion from this work stated above: that finding a ``global'' constant $\etaeff$ or $\zetaeff$ that mimics correctly the temperature dependence of $\eta/s(T)$ or $\zeta/s(T)$ appears only possible for 0+1D fluid.

Nevertheless, depending on the problem under study, it is possible that certain regions of the fluid are more relevant than others: that is, it might be more important for the effective viscosity to be the same along certain characteristics than others. In such cases, one can tailor a definition of approximate global effective viscosity that takes this into account. This must be done on a case-by-case basis, and does not constitute a general definition of global effective viscosity, although it can certainly have practical applications.

\section{Implication for hydrodynamics studies of heavy ion collisions}

\label{sec:hic}

There is evidence from hydrodynamic studies of heavy ion collisions that the bulk viscosity of QCD is difficult to constrain~\cite{Bernhard:2019bmu}: for example, a wide but low peak for $\zeta/s(T)$ cannot easily be distinguished from a narrow but high peak~\cite{Bernhard:2019bmu}.
Studies with shear viscosity, Ref.~\cite{Niemi:2015qia} for example, have also found that different temperature dependence of $\eta/s$ can be difficult to tell apart.
Note that these works do not imply that the viscosities of QCD could not be determined by studying heavy ion collisions: their point is rather that the viscosities can be difficult to extract, and may require a more diverse array of experimental measurements than previously thought.

In this work, we discussed how to identify different parametrizations of $\zeta/s(T)$ and $\eta/s(T)$ that lead to similar hydrodynamic evolution. In the 0+1D case, we found that large families of $\eta/s(T)$ and $\zeta/s(T)$ can lead to almost identical temperature evolutions. In this specific and highly symmetric 0+1D scenario, it is clear that one would not be able to extract the temperature dependence of the viscosities from the temperature profile alone.
In 1+1D, we also identified families of $\eta/s(T)$ and $\zeta/s(T)$ that lead to similar hydrodynamic evolution. However, the transverse dynamics of the system makes it significantly more difficult to find such families of $\eta/s(T)$ and $\zeta/s(T)$.
Moreover, while the temperature profiles obtained were similar, they could presumably be differentiated with sufficiently precise information on the temperature profile.
For fluids with fewer symmetries, it thus appears unlikely that wide classes of $\eta/s(T)$ and $\zeta/s(T)$ could lead to hydrodynamic evolutions difficult to differentiate.

While studies such as Refs.~\cite{Bernhard:2019bmu} and \cite{Niemi:2015qia} motivated this work, it has to be emphasized that it is still early to establish connections between the two. The problem studied in these earlier works is actually different from the one discussed here: in the present work, we investigated families of equivalent viscosities in a given fluid (with a fixed initial temperature profile, for example). In  Refs.~\cite{Bernhard:2019bmu} and \cite{Niemi:2015qia}, the initial temperature profile is allowed to vary when identifying equivalent viscosities. Moreover, the fluid studied in these publications were full 2+1D hydrodynamics, not 0+1D or 1+1D.

Nevertheless, lessons about the concept of effective shear and bulk viscosities can be drawn from this work and can help better understand the study of heavy ion collisions.
We saw in Section~\ref{sec:bjorken_ns} that a constant $\eta/s$ or $\zeta/s$ could generally mimic well a 0+1D fluid with any temperature-dependent viscosity. There should thus rarely ever be a reason to study 0+1D fluids with temperature-dependent viscosities: a simpler and equivalent constant $\eta/s$ or $\zeta/s$ should be  used instead.
On the other hand, we saw in Section~\ref{sec:ns_1_plus_1} that the concept of equivalent constant $\eta/s$ or $\zeta/s$ must be abandoned already at 1+1D: the effective viscosity is dependent on the transverse position in the fluid.
In 2+1D hydrodynamic studies of heavy ion collisions, it is not uncommon to assume that a temperature-dependent $\eta/s(T)$ or $\zeta/s(T)$ can be mimicked by a constant effective viscosity. The present work suggests that such an approximation is unlikely to be particularly precise, and should only be made when the exact details of the hydrodynamic evolution are not too important for the study in question.

\section{Summary and outlook}

In its simplest incarnation, the concept of effective viscosity is a constant value of $\zeta/s$ or $\eta/s$ which produces an equivalent hydrodynamic evolution as that obtained with a temperature-dependent $\zeta/s(T)$ or $\eta/s(T)$. We discussed in Section~\ref{sec:bjorken_ns} that such effective viscosity can indeed be found for a 0+1D boost-invariant system. It is thus generally unnecessary to study these systems with a temperature-dependent viscosity. Moreover it is straightforward to calculate the effective viscosity corresponding to arbitrary parametrizations of $\zeta/s(T)$ or $\eta/s(T)$.
This conclusion is not related to the boost invariance of the system, but rather to the 0+1D nature of the system: as such, we expect that studying 0+1D system with other symmetries (Hubble expansion for example) would lead to similar conclusions.

In Section~\ref{sec:ns_1_plus_1}, we generalized the concept of effective viscosity to 1+1D cylindrically-symmetric boost-invariant fluid. We discussed how to identify different $\zeta/s(T)$ or $\eta/s(T)$ that lead to similar hydrodynamic evolution. We showed however that one must already abandon the concept of a single constant effective value of $\zeta/s$ or $\eta/s$, since the effective viscosity depends on the transverse position.

These 0+1D and 1+1D results were obtained with first-order relativistic viscous Navier-Stokes hydrodynamics. To study heavy ion collisions, second-order relativistic (``Israel-Stewart'') viscous hydrodynamics is used. A preliminary study of the effect of second-order corrections on the concept of effective viscosity was performed for a 0+1D fluid in Section~\ref{sec:is_ns}. The approximate definition of ``second-order effective viscosity'' given by Eq.~\ref{eq:Veff_IS_bjorken_qcd} relates (i) the definition of effective viscosity identified in first-order hydrodynamics, (ii) the initial value of the viscous part of the energy-momentum tensor, and (iii) the relaxation time of the system.

As discussed in Section~\ref{sec:hic}, additional work is still necessary to better understand if the results presented in this work can be related to current phenomenological challenges in constraining the shear and bulk viscosity of QCD. Additionally, while some results in the present work focused on bulk viscosity, there is no fundamental challenges to extending them to shear viscosity. This could be used to better understand the interplay between shear and bulk viscosity, and how this can affect the discussion of effective viscosity presented in Section~\ref{sec:ns_1_plus_1}. As discussed above, it could also be worth investigating applications in  0+1D and 1+1D fluid with symmetries different than the ones explored in this work.

\begin{acknowledgments}
	The authors thank Jonah Bernhard, Lin Dai, Lipei Du, Gabriel Denicol, Ulrich Heinz, Matthew Luzum, Aleksas Mazeliauskas, Scott Moreland, Bjoern Schenke, Chun Shen and Derek Teaney for discussions that led to this work and helped improve it. \mbox{J-.F.P.} thanks the organizers and participants of the  Seminar in Hadronic Physics at McGill and of the Nuclear Theory/RIKEN Seminar in BNL for early feedback on this work. This work was supported by
	the U.S. Department of Energy under Award Numbers DE-FG02-05ER41367.
\end{acknowledgments}

\appendix

\section{Solving relativistic ideal hydrodynamics with Bjorken symmetries}

\label{sec:appendix_bjorken}

The ideal hydrodynamic equation for temperature in terms of the variables $\lambda\equiv\ln(\tau/\tau_0)$ and $f(\lambda)\equiv \ln(T_0/T)$ is
\begin{equation}
\partial_\lambda f(\lambda)=c_s^2(f(\lambda)) %
\label{eq:NS_T_conserv_Bj_appendix}
\end{equation}
with $f(\lambda=0)=0$. 
Given $\tau \ge \tau_0$, $\lambda\ge 0$ and $f(\lambda) \ge 0$.

The ideal, conformal case  with $c_s^2=1/3$ reduces to $f(\lambda)=\lambda/3$, which is terms of $T$ and $\tau$ is the well-known
$$
T_{id}^{c}(\tau)=T_0 \left( \frac{\tau_0}{\tau} \right)^{1/3} \; .
$$

In the case of a non-conformal fluid, a good approximate solution can be found if $c_s^2(T)$ 
does not vary too much with temperature, as is the case of QCD. 
Equation~\ref{eq:NS_T_conserv_Bj_appendix} can be written
\begin{equation}
\lambda=\int_{0}^{f(\lambda)} d f^\prime c_s^{-2}(f^\prime) \; .
\label{eq:ideal_f_lambda}
\end{equation}

For QCD, $c_s^{-2}$ varies from $3$ to $\sim 7$ in the range of temperature relevant in collisions at the RHIC and the LHC (c.f. Fig.~\ref{fig:cs2}). We first write an approximate solution $\bar{f}(\lambda)$ with an effective speed of sound $c_s^{-2}(f^\prime)=\bar{c}_s^{-2}$:
\begin{equation}
\bar{f}(\lambda) \equiv \bar{c}_s^{2} \lambda 
\end{equation}
which for $\bar{c}_s^{-2}\sim 3-7$ should be a reasonable first approximation of $f(\lambda)$. Using $\bar{f}(\lambda)$, Eq.~\ref{eq:ideal_f_lambda} can be written
\begin{equation}
\lambda=\left[ \int_{0}^{\bar{f}(\lambda)} d f^\prime c_s^{-2}(f^\prime) \right]+\left[ \int_{\bar{f}(\lambda)}^{f(\lambda)} d f^\prime c_s^{-2}(f^\prime) \right] \; .
\end{equation}
The first term can be calculated ``exactly'' by integrating the speed of sound. The second term can be calculated by expanding $c_s^{-2}(f^\prime)$ as a power series:
\begin{equation}
c_s^{-2}(f^\prime)=c_s^{-2}(\bar{f})+\left.\frac{d c_s^{-2}}{d f}\right|_{f=\bar{f}} (f-\bar{f})+\ldots
\label{eq:csm2_exp}
\end{equation}

Truncating at the first term, we obtain
\begin{equation}
\lambda\approx \left[ \int_{0}^{\bar{f}(\lambda)} d f^\prime c_s^{-2}(f^\prime) \right] +\left[ c_s^{-2}(\bar{f}(\lambda)) \left(f(\lambda)-\bar{f}(\lambda)\right) \right] \; .
\end{equation}
This gives
\begin{equation}
f(\lambda)\approx \bar{f}(\lambda)+ c_s^{2}(\bar{f}(\lambda)) \left[\lambda - \int_{0}^{\bar{f}(\lambda)} d f^\prime c_s^{-2}(f^\prime) \right] 
\end{equation}
which can be rewritten
\begin{equation}
f(\lambda)\approx \bar{f}(\lambda)- \left[ \int_{0}^{\bar{f}(\lambda)} d f^\prime \frac{c_s^{-2}(f^\prime) - \bar{c}_s^{-2}}{c_s^{-2}(\bar{f}(\lambda))} \right] \; .
\end{equation}
In general the speed of sound is a slowly varying function over the range of integration, and the formula above can be simplified further by evaluating the integrand at $f^\prime=\bar{f}(\lambda)$:
\begin{equation}
f(\lambda)\approx \bar{f}(\lambda) \left[1- \frac{c_s^{-2}(\bar{f}(\lambda)/2) - \bar{c}_s^{-2}}{c_s^{-2}(\bar{f}(\lambda))} \right] \; .
\label{eq:Bjoerken_ideal_f_lambda_int}
\end{equation}

In terms of temperature, this solution can be written
\begin{equation}
T_{id}(\tau)=T_0 \left[ \frac{\bar{T}_{id}(\tau)}{T_0} \right]^{\left(1- \frac{c_s^{-2}(\sqrt{\bar{T}_{id}(\tau) T_0}) - \bar{c}_s^{-2}}{c_s^{-2}(\bar{T}_{id}(\tau))} \right) } 
\label{eq:bjorken_ideal_approx}
\end{equation}
with
$$
\bar{T}_{id}(\tau)=T_0 \left( \frac{\tau_0}{\tau} \right)^{\bar{c}_s^{2}} \; .
$$

\begin{figure}[tb]
	\centering
	\includegraphics[width=0.5\textwidth]{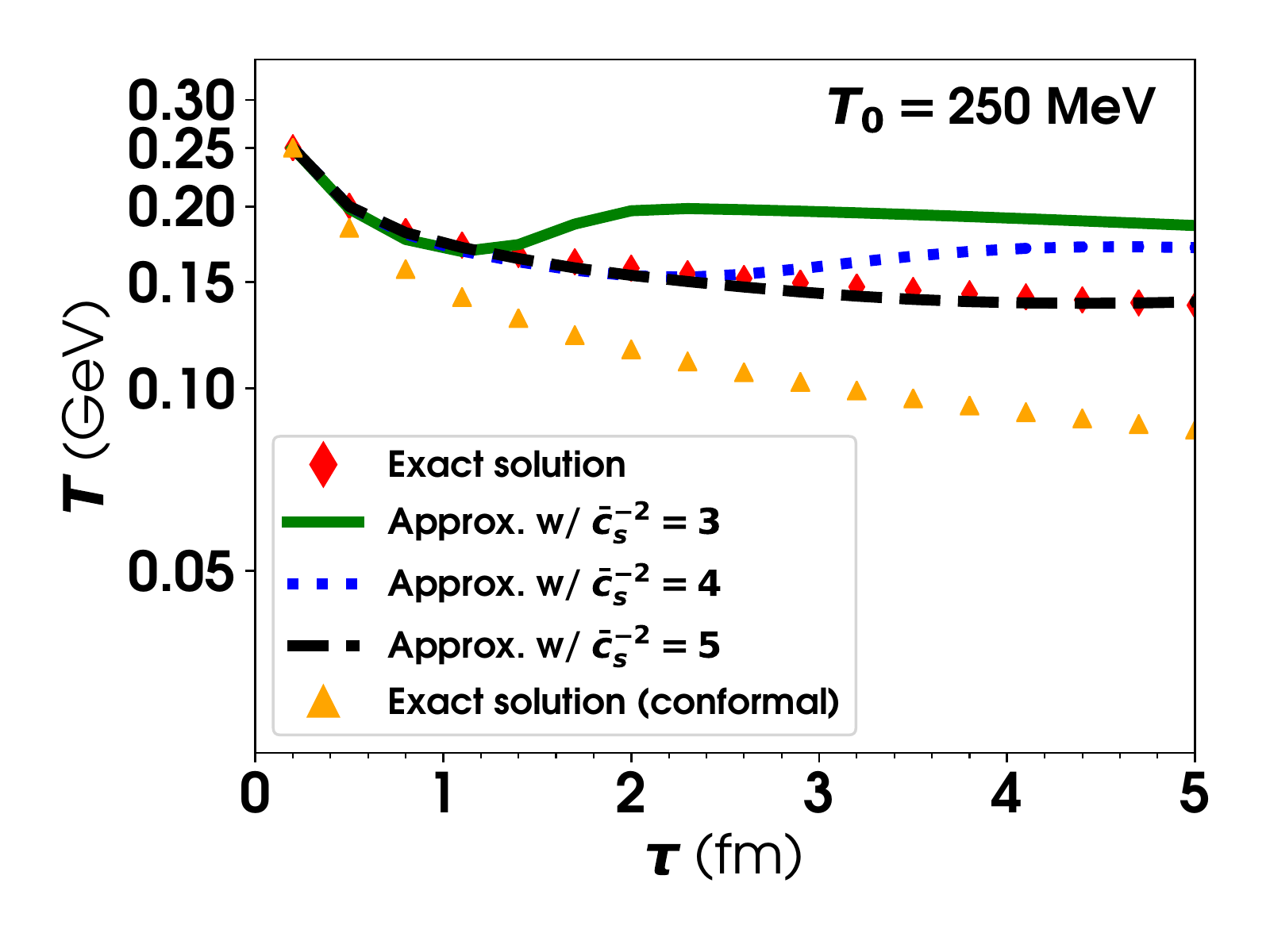}
	(a)
	\includegraphics[width=0.5\textwidth]{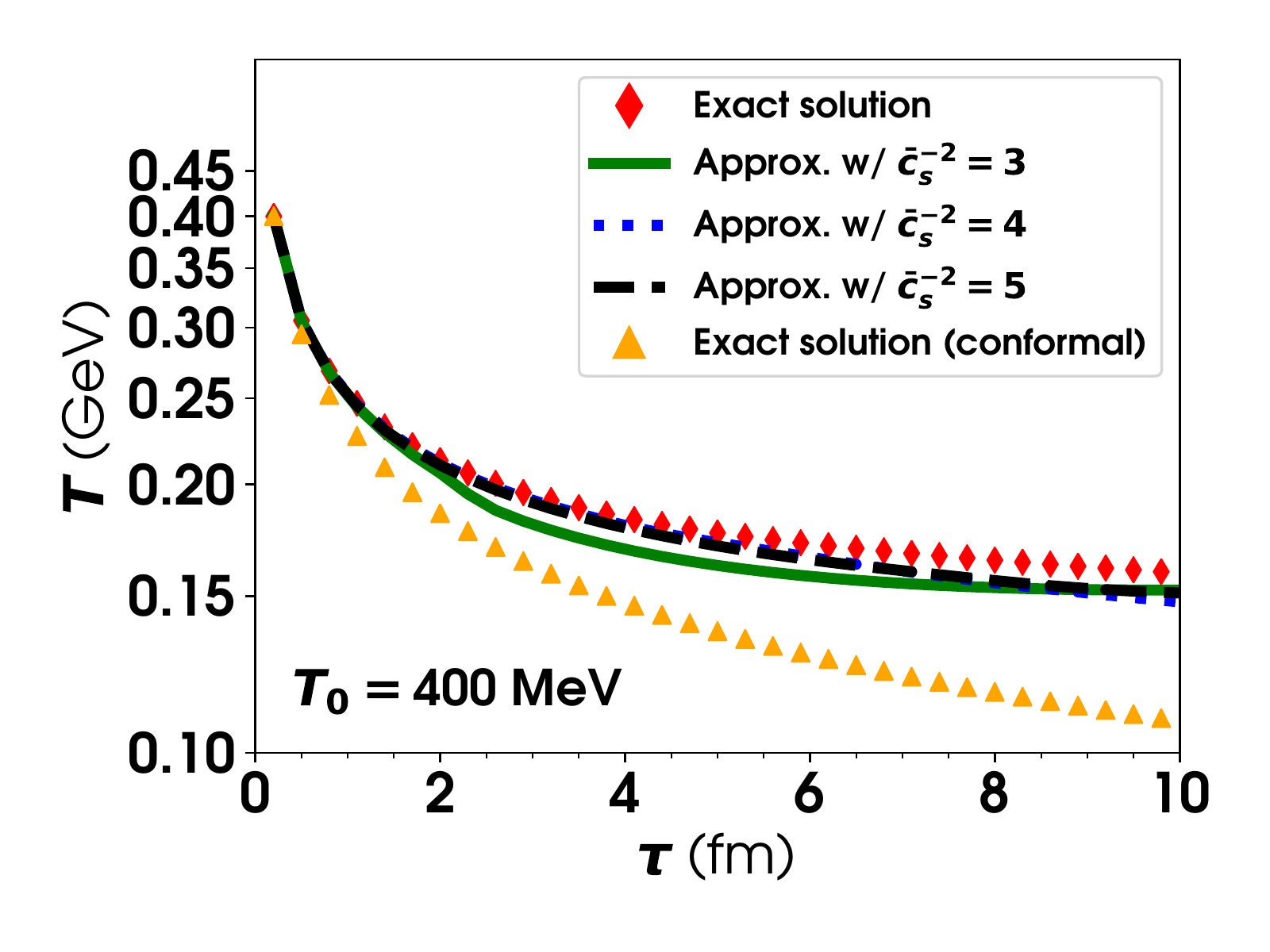}
	(b)
	\caption{Temperature evolution as a function of $\tau$ given by the approximate solution to ideal Bjorken hydrodynamics, Eq.~\ref{eq:bjorken_ideal_approx}, for $\bar{c}_s^{-2}=3,4,5$, compared with the exact solution.  
		The exact conformal solution is shown for reference. For (a) $T_0=250$~MeV and (b) $T_0=400$~MeV.}
	\label{fig:ideal_bjorken_approx}
\end{figure}

Although the choice $\bar{c}_s^{-2}=3$ (the well-known conformal solution) appears natural, there is a good reason to choose a larger value. Equation~\ref{eq:csm2_exp} was written explicitly, although only the first term was kept, to highlight that the radius of convergence is related to the difference $[f-\bar{f}=\ln(\bar{T}/T)]$. For the QCD equation of state, choosing $\bar{c}_s^{-2}=3$ will result in a good solution at large $T_0$ and small $\tau$ but may produce a divergent solution at larger $\tau$. A value of $\bar{c}_s^{-2} \sim 4-5$ sidestep this issue for any range of temperatures relevant for heavy ion collisions. This is illustrated on Fig.~\ref{fig:ideal_bjorken_approx}.

\section{Effective viscosity in boost-invariant Navier-Stokes hydrodynamics: a different approach to effective viscosities}

\label{sec:effective_visc_Bjorken_optimization}

In the Bjorken 0+1D case, the particlization hypersurface is a surface at a fixed time $\tau=\tau_\textrm{particliz}$. This time can be found by first writing the Bjorken solution $\tau(T)$ instead of the usual $T(\tau)$, and then solve  $\tau_\textrm{particliz}=\tau(T_{\textrm{particliz}})$. Finally, we must find $\langle \eta/s \rangle_{\textrm{eff}}$ and $\langle \zeta/s \rangle_{\textrm{eff}}$ that minimizes:
\begin{equation}
\left[ \tau^*(T_\textrm{particliz})- \tau(T_{\textrm{particliz}}) \right]^2
\label{eq:Bjorken_optimization}
\end{equation}
where $\tau^*(T)$ is the solution with effective viscosities.

Equation~\ref{eq:NS_T_conserv_Bj} for the evolution of temperature in a Bjorken system can be rewritten
\begin{eqnarray}
\frac{\partial \ln\tau}{\partial \ln T}&=&-c_s^{-2}(T) \left[ 1-\frac{V(T)}{\tau T}  \right]^{-1} \nn \\
& \approx & -c_s^{-2}(T) \left[ 1+\frac{V(T)}{\tau T}  \right] \; .
\label{eq:NS_tau_conserv_Bj}
\end{eqnarray}
For the second equation, we used the fact that viscous corrections are modest to simplify the right-hand side of the equation.

The expression for $\tau(T)$ is thus
\begin{equation}
\ln\left(\frac{\tau}{\tau_0}\right)\approx \int_{T}^{T_0} \frac{d T^\prime}{T^\prime} c_s^{-2}(T^\prime) \left[ 1+\frac{V(T^\prime)}{\tau T^\prime}  \right]
\end{equation}
which can be solved iteratively, first without viscous effects (ideal case):
\begin{equation}
\ln\left(\frac{\tau_{(I)}}{\tau_0}\right)= \int_{T}^{T_0} \frac{d T^\prime}{T^\prime} c_s^{-2}(T^\prime) 
\end{equation}
and then with viscous corrections:
\begin{eqnarray}
\ln\left(\frac{\tau_{(II)}}{\tau_0}\right)&=&  \int_{T}^{T_0} \frac{d T^\prime}{T^\prime} c_s^{-2}(T^\prime) \left[ 1+\frac{V(T^\prime)}{\tau_{(I)} T^\prime}  \right] \nn \\
&=&  \ln\left(\frac{\tau_{(I)}}{\tau_0}\right) + \Delta_{\tau_{(II)}} \nn \\
\end{eqnarray}
with
\begin{equation}
\Delta_{\tau_{(II)}} \equiv \int_{T}^{T_0} \frac{d T^\prime}{T^\prime} c_s^{-2}(T^\prime) \frac{V(T^\prime)}{\tau_{(I)} T^\prime} \; .
\end{equation}

The interpretation of the positive-definite $\Delta_{\tau_{(II)}}$ is that viscosity generates entropy, and consequently the temperature decreases more slowly in the viscous case than the ideal one. This implies that it takes longer to reach a given particlization temperature.
The ratio between the particlization time in the viscous and ideal cases is
\begin{equation}
\frac{\tau_{viscous}}{\tau_{ideal}}\approx \frac{\tau_{(II)}}{\tau_{(I)}} = \exp\left(  \Delta_{\tau_{(II)}}  \right ) 
\; .
\end{equation}

As discussed in Section~\ref{sec:ns_bjorken_qcd} (Eq.~\ref{eq:tau_vs_T_ideal}), $\tau_{(I)}$ can be approximated as
\begin{equation}
\tau_{(I)}\approx \tau_0 \left( \frac{T_0}{T} \right)^{c_s^{-2}(\sqrt{T T_0})}
\label{eq:tauI_appox}
\end{equation}
meaning that
\begin{equation}
\frac{\tau_{viscous}}{\tau_{ideal}}\approx \exp \! \left[ \frac{1}{\tau_0} \int_{T}^{T_0} \!\!\! \frac{d T^\prime}{(T^\prime)^2} c_s^{-2}(T^\prime) V(T^\prime) \left( \frac{T^\prime}{T_0} \right)^{c_s^{-2}(\sqrt{T^\prime T_0})} \right]
\label{eq:appendix_tau_visc_bjorken_estimate}
\end{equation}

Assuming\footnote{Assuming an average speed of sound $\bar{c}_s$, the weight in the exponential is $\bar{c}_s^2/(\bar{c}_s^2-1)$, which is $3/2$ for $c_s^{-2}=3$ and $5/4$ for $c_s^{-2}=5$. The breaking of conformality in QCD ($c_s^{-2} \gtrsim 3$, see Fig.~\ref{fig:cs2}) thus reduces Eq.~\ref{eq:tau_visc_vs_ideal} by $\sim 10$\%.} $c_s^{-2} \approx 3$ and with a constant viscosity,
\begin{eqnarray}
\frac{\tau_{viscous}}{\tau_{ideal}}&\approx& \exp \left[\frac{3 V_{\textrm{eff}}}{\tau_0}  \int_{T}^{T_0} \frac{d T^\prime}{T^\prime} \frac{1}{T^\prime} \left( \frac{T_0}{T^\prime} \right)^{-3} \right] \nn \\
& \overset{T\ll T_0}{\approx} & \exp \left[\frac{3 V_{\textrm{eff}}}{2 \tau_0 T_0} \right] 
\label{eq:tau_visc_vs_ideal}
\end{eqnarray}
with $V_{\textrm{eff}}$ the effective viscosity.
The equation above provides a simple estimate for the effect of viscosity on the lifetime of a Bjorken system. 

Alternatively, a simple estimate of the effective viscosity can be obtained by knowing $\tau_{viscous}$ and $\tau_{ideal}$:
\begin{equation}
V_{\textrm{eff}} \approx \frac{2}{3} \tau_0 T_0 \left( \frac{\tau_{viscous}}{\tau_{ideal}} -1 \right) \; .
\end{equation}

Returning to Eq.~\ref{eq:appendix_tau_visc_bjorken_estimate}, the particlization time can be estimated with $\tau_\textrm{particliz}\equiv \tau_{viscous} \approx \tau_{(II)}$.

Equation~\ref{eq:Bjorken_optimization} can thus be written
\begin{eqnarray}
M& \equiv &\left[ \tau^*(T_\textrm{particliz})- \tau(T_{\textrm{particliz}}) \right]^2 \nn \\
& \approx & \left[ \tau_0 \exp\left(\int_{T}^{T_0} \frac{d T^\prime}{T^\prime} c_s^{-2}(T^\prime) \left[ 1+\frac{V_{\textrm{eff}}}{\tau_{(I)} T^\prime}  \right]\right) - \right. \nn \\
& & \left. \tau_0 \exp\left(\int_{T}^{T_0} \frac{d T^\prime}{T^\prime} c_s^{-2}(T^\prime) \left[ 1+\frac{V(T^\prime)}{\tau_{(I)} T^\prime}  \right]\right) \right]^2 \nn \\
& \approx &  \left[ \tau_0 \exp\left(\int_{T}^{T_0} \frac{d T^\prime}{T^\prime} c_s^{-2}(T^\prime)\right) \right]^2 \nn \\
& & \left[ \exp\left(\int_{T}^{T_0} \frac{d T^\prime}{T^\prime} \frac{c_s^{-2}(T^\prime) V_{\textrm{eff}}}{\tau_{(I)} T^\prime} \right) \right. \nn \\
& & - \left. \exp \left( \int_{T}^{T_0} \frac{d T^\prime}{T^\prime} \frac{c_s^{-2}(T^\prime) V(T^\prime)}{\tau_{(I)} T^\prime} \right) \right]^2 \; . \nn \\
\end{eqnarray}

Optimizing with respect to $V_{\textrm{eff}}$, $\partial M/\partial V_{\textrm{eff}} \equiv 0$, yields
\begin{multline}
\left[ \exp \left( \int_{T}^{T_0} \frac{d T^\prime}{T^\prime} \frac{c_s^{-2}(T^\prime) V_{\textrm{eff}}}{\tau_{(I)} T^\prime} \right) \right. \nn \\ 
\left. - \exp \left( \int_{T}^{T_0} \frac{d T^\prime}{T^\prime} \frac{c_s^{-2}(T^\prime) V(T^\prime)}{\tau_{(I)} T^\prime} \right) \right] \\
\times \exp \left[ \int_{T}^{T_0} \frac{d T^\prime}{T^\prime} \frac{c_s^{-2}(T^\prime) V_{\textrm{eff}}}{\tau_{(I)} T^\prime} \right] \int_{T}^{T_0} \frac{d T^\prime}{T^\prime} \frac{c_s^{-2}(T^\prime)}{\tau_{(I)} T^\prime}\equiv 0 \; .
\end{multline}
The second line is positive-definite, implying the expected result
\begin{multline}
\exp \left( \int_{T}^{T_0} \frac{d T^\prime}{T^\prime} \frac{c_s^{-2}(T^\prime) V_{\textrm{eff}}}{\tau_{(I)} T^\prime} \right) = \\
 \exp \left( \int_{T}^{T_0} \frac{d T^\prime}{T^\prime} \frac{c_s^{-2}(T^\prime) V(T^\prime)}{\tau_{(I)} T^\prime} \right) \; ,
\end{multline}
\begin{equation}
\Rightarrow V_{\textrm{eff}} = \frac{\int_{T}^{T_0} \frac{d T^\prime}{T^\prime} \frac{c_s^{-2}(T^\prime) V(T^\prime)}{\tau_{(I)} T^\prime}}{\int_{T}^{T_0} \frac{d T^\prime}{T^\prime} \frac{c_s^{-2}(T^\prime) }{\tau_{(I)} T^\prime}} \; .
\end{equation}

Using Eq.~\ref{eq:tauI_appox} as approximation for $\tau_{(I)}$, we get
\begin{equation}
V_{\textrm{eff}} = \frac{\int_{T}^{T_0} d T^\prime \left( \frac{T^\prime}{T_0} \right)^{c_s^{-2}(\sqrt{T^\prime T_0})-2} c_s^{-2}(T^\prime) V(T^\prime)}{\int_{T}^{T_0} d T^\prime \left( \frac{T^\prime}{T_0} \right)^{c_s^{-2}(\sqrt{T^\prime T_0})-2} c_s^{-2}(T^\prime)} \; .
\label{eq:V_eff_Bj_from_hypersurface}
\end{equation}

Equation~\ref{eq:V_eff_Bj_from_hypersurface} differs from the previously derived equation for the effective viscosity, Eq.~\ref{eq:Veff_bjorken_qcd}, by a factor $c_s^{-2}(T^\prime)$. This factor originates ultimately from comparing times ($\tau$) instead of temperature to find the optimal effective viscosity $V_{\textrm{eff}}$, since $d \ln(T) \sim c_s^2(T) d \ln(\tau)$. In practice, Eqs~\ref{eq:Veff_bjorken_qcd} and \ref{eq:V_eff_Bj_from_hypersurface} generally give similar results, since $c_s^{-2}(T^\prime)$ is a slowly varying function, an assumption that we used to obtain both equations. We verified, for example, that the results obtained in Sections~\ref{sec:ns_bjorken_qcd_shear} and \ref{sec:ns_bjorken_qcd_bulk}  do not change significantly when calculated from Eq.~\ref{eq:V_eff_Bj_from_hypersurface} rather than Eq.~\ref{eq:Veff_bjorken_qcd}.

\bibliographystyle{ieeetr}
\bibliography{biblio}

\end{document}